\documentclass[review]{elsarticle}
\pdfoutput=1

\usepackage{lineno,hyperref}
\modulolinenumbers[5]

\journal{Journal of \LaTeX\ Templates}

\usepackage{numcompress}

\usepackage{float}
\usepackage{float}
\usepackage{amsmath}
\usepackage{amssymb}
\usepackage[dvipsnames]{xcolor}
\usepackage{enumerate}
\usepackage[]{algorithm2e}
\usepackage{physics,soul}

\newcommand{\Nfft}{N_{\mathrm{fft}}}
\newcommand{\Nblk}{N_{\mathrm{blk}}}

\newcommand{\Novlp}{N_{\mathrm{ovlp}}}
\newcommand{\Nwin}{N_{\mathrm{win}}}
\newcommand{\Ndof}{N_{\mathrm{dof}}}
\newcommand{\bw}{b_\mathrm{win}}
\newcommand{\Nsample}{M}
\newcommand{\df}{\delta\!f}

\newcommand{\eqn}[1]{(\ref{#1})}
\DeclareMathSymbol{\shortminus}{\mathbin}{AMSa}{"39}

\newcommand{\new}[1]{{#1}}
\newcommand{\ri}[1]{{#1}}
\newcommand{\rii}[1]{{#1}}

\begin{document}

\begin{frontmatter}

\title{Spectral Proper Orthogonal Decomposition using Multitaper Estimates}

\author{Oliver T. Schmidt\fnref{myfootnote}}
\address{Department of Mechanical and Aerospace Engineering, University of California San Diego, La Jolla, CA 92093, USA}
\fntext[myfootnote]{oschmidt@ucsd.edu}

%
%

\begin{abstract}

The use of multitaper estimates for spectral proper orthogonal decomposition (SPOD) is explored. Multitaper and multitaper-Welch estimators that use discrete prolate spheroidal sequences (DPSS) as orthogonal data windows are compared to the standard SPOD algorithm that exclusively relies on weighted overlapped segment averaging, or Welch's method, to estimate the cross-spectral density matrix. Two sets of turbulent flow data, one experimental and the other numerical, are used to discuss the choice of resolution bandwidth and the bias-variance tradeoff. Multitaper-Welch estimators that combine both approaches by applying orthogonal tapers to overlapping segments allow for flexible control of resolution, variance, and bias. At additional computational cost but for the same data, Multitaper-Welch estimators provide lower variance estimates at fixed frequency resolution or higher frequency resolution at similar variance compared to the standard algorithm.


\end{abstract}


\end{frontmatter}

\linenumbers

\section{Introduction}

SPOD is the specialization of the most general form of proper orthogonal decomposition (POD, \cite{Lumley:1970}) to statistically stationary data and is best computed in the frequency domain. An early application of the method can be found in the experimental work of \citet{glauser1987coherent}. The method has recently gained popularity as theoretical connections to dynamic mode decomposition (DMD,\cite{schmid2010dmd}) and resolvent analysis of turbulent mean flows \citep{McKeonSharma2010} were established by \citet{towneschmidtcolonius_2018_jfm}. Another factor is its application to large numerical data that were specifically generated for this purpose \citep{SchmidtEtAl_2018_JFM}. SPOD has been applied to a broad range of canonical \cite{arndt1997proper,CitrinitiGeorge2000,nidhanetal_2020_prf,gordeyev2000coherent,gudmundsson2011instability,hellstrom2014,tutkun2017lumley} and technical \cite{ArayaEtAl_2017_JFM,abreu2017coherent,he2021spectral,li2021research} turbulent flows, weather and climate data \cite{schmidtetal_2019_mwr}, and found diverse applications in aeroacoustics \cite{sanjose2019modal,nekkantischmidt_2020_aiaaj}, stochastic estimation \cite{baars2014proper}, frequency-time analysis \cite{nekkantischmidt_jfm_2021}, flow field reconstruction \cite{ghate_towne_lele_2020,nekkantischmidt_jfm_2021}, and reduced-order modeling \cite{chuschmidt_2021_tcfd,towne2021space}.

The SPOD modes and their mode energies are, respectively, the eigenvectors and eigenvalues of the weighted cross-spectral density (CSD) matrix. \rii{The CSD matrix is estimated from an ensemble of realizations of the temporal discrete Fourier transform (DFT). In contrast to classical Fourier analysis (see, e.g., \citet{pain2019large} for a recent application to large data), SPOD modes are, in an inner product norm of choice, optimal linear combinations of many Fourier modes.} The standard approach to estimate the CSD matrix for SPOD is weighted overlapped segment averaging (WOSA), or Welch's method \citep{welch1967use,schmidtcolonius_2020_aiaaj}. A different approach that is routinely used for the estimation of power spectra from time signals is the multitaper (MT), or Thomson's multitaper method \citep{thomson1982spectrum}. The multitaper method corresponds to weighted averaging of independent spectral estimates that are obtained from multiple orthogonal data windows. Most notably, it has been demonstrated by \citet{bronez1992performance} that the multitaper power spectrum estimator outperforms the Welch estimator in terms of leakage, variance, and resolution if two of these three quantities are fixed to determine the third. In the context of spectral analysis of turbulent flow data, a frequency–wavenumber analysis technique that uses multitaper estimates was proposed by \citet{geoga_haley_siegel_anitescu_2018}.

The remainder of the paper is organized as follows. The technical background is discussed in \S \ref{sec:methods} and includes a brief summary of the discrete SPOD problem
(\S \ref{sec:spod}) and DPSS (\S \ref{sec:slepian}), as well as a recapitulation of Welch's method, and the introduction of multitaper estimates for SPOD (\S \ref{sec:estimates}). The two example data sets are presented in \S \ref{sec:data}. In \S \ref{sec:results}, SPOD analyses conducted using the basic Welch and single-block multitaper estimates (\S \ref{sec:results_intro}) are compared and discussed in the light of parameter selection, frequency resolution, bandwidth, and the variance-bias tradeoff. Next, multitaper-Welch estimates that combine the advantages of both methods are computed for the two examples and discussed in \S \ref{sec:results_main}. \new{The additional computational cost associated with multitaper estimates is addressed in \S\ref{sec:performance}.} Finally, implications and future applications of this work are discussed in \S \ref{sec:discussion}.

\section{Background and Methodology}\label{sec:methods}

\subsection{Spectral proper orthogonal decomposition (SPOD)}\label{sec:spod}

The discrete SPOD of an ensemble of ${N_t}$ snapshots of spatio-temporal multivariate data with zero mean,\ri{
\begin{eqnarray}\label{eqn:data}
\vb{q}_i=\vb{q}(t_i),\quad i=1,\cdots,{N_t},\quad \vb{q}\in\mathbb{R}^{\Ndof \times 1}
\end{eqnarray}
 here written as a sequence of column vectors, is obtained from the eigendecompositions
\begin{equation}\label{eqn:evp}
\vb{C}_j\vb{W}\vb{\Phi}_j=\vb{\Phi}_j\vb{\Lambda}_j,\quad j=1,\cdots,\Nfft,\quad \vb{C}_j\in\mathbb{R}^{\Ndof \times \Ndof},
\end{equation}
of the cross-spectral density (CSD) matrices, $\vb{C}_j$. The rank $\Nsample$ of the CSD matrices corresponds to the number of independent realizations of the Fourier transform that are used for their construction, and their size is determined by the number of degrees of freedom (number of variables times number of points in space), $\Ndof$.
The number of eigenvalue problems to be solved corresponds to the number of discrete frequencies, $\Nfft$, of the discrete Fourier transform in the permissible frequency band between zero and the Nyquist frequency, $f_\mathrm{Nyq}=\frac{f_s}{2}$, where $f_s=1/\Delta t$ is the sampling frequency.} The columns of $\vb{\Phi}_j=\qty[\boldsymbol{\phi}_{j}^{(1)}, {\boldsymbol{\phi}}_{j}^{(2)},\dots,{\boldsymbol{\phi}}_{j}^{(\Nsample)}]$, i.e., the eigenvectors of the weighted CSD matrices $\vb{C}_j\vb{W}$, are the SPOD modes and the diagonal entries of $\vb{\Lambda}_j$, i.e., the corresponding eigenvalues $\lambda_f^{(1)}\geq\lambda_f^{(2)}\geq\dots\geq\lambda_f^{(\Nsample)}$, the mode energies. The weight matrix $\vb{W}$ accommodates spatial integration and, for multivariate data, variable-dependent weights. The discrete (sample) spatial CSD matrix is given as
\begin{subequations}
\begin{align}\label{eqn:csd}
\vb{C}_j &= \frac{1}{\Nsample}\sum_{k=1}^{\Nsample} \hat{\vb{q}}_{j}^{(k)} \qty(\hat{\vb{q}}_{j}^{(k)})^*\\ &= \frac{1}{\Nsample}\hat{\vb{Q}}_{j}  \hat{\vb{Q}}_{j}^{*},\label{eqn:csd_mat}
\end{align}
\end{subequations} 
where $\hat{\vb{q}}_{j}^{(k)}$ is the $k$-th out of $\Nsample$ realizations of the discrete Fourier transform at the $j$-th frequency. The matrix representation, equation (\ref{eqn:csd_mat}), uses the data matrix $\hat{\vb{Q}}_{j}=\qty[\hat{\vb{q}}_{j}^{(1)}, \hat{\vb{q}}_{j}^{(2)},\dots,\hat{\vb{q}}_{j}^{(\Nsample)}]$. The denominator is either $\Nsample$ or $\Nsample-1$, depending on whether the true mean, if available, or the sample mean is subtracted from the data, respectively. \ri{
In the common case where $\Nsample < \Ndof$, the smaller size $\Nsample \times \Nsample$ eigenvalue problem,
\begin{subequations}
\begin{align}
\label{eqn:evp_econ}
\vb{C}'_j&\vb{\Psi}_j=\vb{\Psi}_j\vb{\Lambda}'_j,\quad \vb{\Phi}_j=\hat{\vb{Q}}_{j}\vb{\Psi}_j,\quad j=1,\cdots,\Nfft,\\
\mathrm{where}\;\vb{C}'_j&=\frac{1}{\Nsample}\hat{\vb{Q}}_{j}^{*}\vb{W}\hat{\vb{Q}}_{j},\quad \vb{C}'_j\in\mathbb{R}^{\Nsample \times \Nsample},\label{eqn:C_econ}
\end{align}
\end{subequations} 
is the weighted temporal CSD matrix and $\vb{\Lambda}'_j\in\mathbb{R}^{\Nsample \times \Nsample}$ the reduced matrix of non-zero eigenvalues, can be solved for the SPOD expansion coefficients, $\vb{\Psi}_j$, instead of using equation \eqn{eqn:evp} to obtain the SPOD modes directly.} The desirable mathematical properties of the SPOD, its continuous formulation, and algorithmic implementations are discussed in detail elsewhere \citep{towneschmidtcolonius_2018_jfm,schmidtcolonius_2020_aiaaj}. 

\subsection{Discrete Prolate Spheroidal Sequences (DPSS)}\label{sec:slepian}

\begin{figure}[H]
  \includegraphics[width=1\textwidth,trim=0 0cm 0cm 0cm,clip]{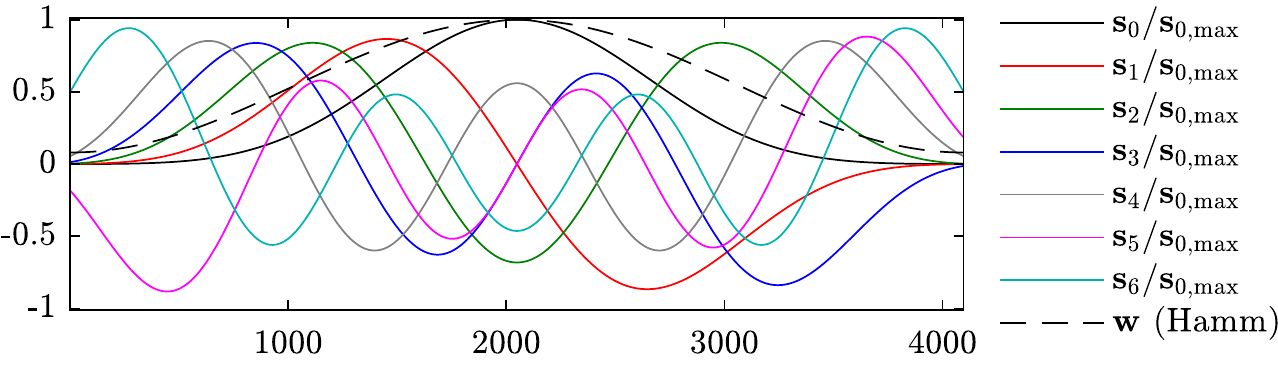}
\caption{Leading seven DPSS for $N=4096$ and $\bw=4$ compared to a standard Hamming window. The DPSS are normalized by the maximum value of $\vb{s}_1$ for clarity.}
\label{fig:slepian}       
\end{figure}

Discrete prolate spheroidal sequences (DPSS), or Slepians, are the discrete-time equivalent of prolate spheroidal wave functions \cite{slepian1961prolate,slepian1978prolate}. They are the optimal solutions to the spectral concentration problem that seeks orthogonal time sequences whose discrete Fourier transform is maximally localized in a frequency band $[-\df,\df]$, defined by the resolution half-bandwidth, $\df$. Given the window length, $N$, and resolution half-bandwidth, $\df$, a common way to find the DPSS is the eigendecomposition
\begin{equation}\label{eqn:slepian}
\vb{H}\vb{s}=\gamma\vb{s},
\end{equation}
where the entries of the Hermitian matrix $\vb{H}=\vb{H}(\df,N)$ are given by
\begin{equation}
h_{ij} = \frac{\sin\qty(2\pi \df (j-i))}{\pi(j-i)},\quad i,j=0,\dots,N-1.
\end{equation}
The eigenvalues $1>\gamma_0>\gamma_1>\dots>\gamma_{N-1}>0$ correspond to the spectral concentrations of energy of the respective tapers, $\vb{s}_0,\vb{s}_1,\dots,\vb{s}_{N-1}$, in the frequency band $[-\df,\df]$. Since $\vb{H}$ is Hermitian, the DPSS are mutually orthogonal. It can be shown that only $2\df N$ (Shannon number) of the spectral concentrations are close to unity. It is hence customary to use only the first $\Nwin=2\df N-1$ Slepian tapers as windows. In terms of the time-halfbandwidth product, $\bw=\df N$, this statement becomes
\begin{equation}\label{eqn:nwin}
	\Nwin=\lfloor 2\bw \rfloor-1,
\end{equation}
and guarantees that the Fourier transforms of all windows are well-concentrated in $[-\df,\df]$ to minimize spectral leakage to neighboring frequencies. The leading seven DPSS for $N =4096$ and $\bw=4$ are compared to a standard Hamming window in figure \ref{fig:slepian}.

\subsection{Welch and multitaper estimates of the SPOD}\label{sec:estimates}

The $N$ independent realizations of the discrete Fourier transform that form the columns of the sample CSD matrix in equation (\ref{eqn:csd}) are obtained in one of two ways. The first option is to obtain well-separated sequences, or blocks, $\vb{q}_{j}^{(k)}$ with $j=1,\dots,\Nfft$ and $k=1,\dots,\Nblk$, one at a time. Denote by $\Nblk$ the total number of blocks, each consisting of $\Nfft$ consecutive snapshots, equally spaced in time by $\Delta t$. This scenario occurs when data acquisition at a high sampling rate is the bottleneck, as is the case in many experimental settings. The second possibility is that the data is acquired as a single, long time series of $N_t$ snapshots, $\vb{q}_j=\vb{q}(t_j)$, with $j=1,\cdots,{N_t}$. This is often the case in numerical settings, where the overall runtime is the bottleneck. In this case, the long time series is segmented into $\Nblk$ blocks, $\vb{q}_{j}^{(k)}=\vb{q}_{j+(k-1)(\Nfft-\Novlp)+1}$ with $j=1,\dots,\Nfft$ and $k=1,\dots,\Nblk$, under the ergodicity hypothesis. Adjacent blocks are allowed to overlap by $\Novlp$ snapshots, which results in a total number of $\Nblk=\left\lfloor \frac{N_t-\Novlp}{\Nfft-\Novlp}\right\rfloor$ blocks. This segmentation is the idea behind the method of \citet{welch1967use}, that was originally devised for the estimation of power spectra of time signals. In the same work, Welch established from theoretical considerations the best practice of using an overlap of 50\%. We follow this best practice but note that the optimal overlap depends on the window function, see \citet{heinzel2002spectrum} for a comprehensive discussion. After segmenting the data, $\Nblk$ realizations of the Fourier transform are obtained from the weighted temporal discrete Fourier transform of each block,
\begin{equation}
\hat{\vb{q}}_{j}^{(k)}=\sum_{i=0}^{\Nfft-1}w[i]\vb{q}_{i}^{(k)}\mathrm{e}^{\frac{\mathrm{i}2\pi}{\Nfft}ij}, \quad k=1,\dots,\Nblk,
\label{eqn:fft}
\end{equation}
where $\hat{\vb{q}}_j^{(k)}$ is the $k$-th realization of the Fourier transform at the $j$-th discrete frequency. It is common practice to use a data window, $\vb{w}$, to prevent spectral leakage due to the non-periodicity of the samples. A popular choice is the symmetric Hamming window,
\begin{equation}
w[i]=0.54-0.46 \cos{\pqty{\frac{2 \pi i}{\Nfft-1} }},\quad i=0, \cdots, \Nfft-1.
\label{eqn:hamm}
\end{equation}
Given the ensemble of realizations, equation (\ref{eqn:fft}), the CSD matrix is formed according to equation (\ref{eqn:csd}), and its eigendecomposition, equation (\ref{eqn:evp}), concludes the standard SPOD algorithm. 

Inspired by the method of \citet{thomson1982spectrum}, also originally intended for the estimation of power spectra, we propose to use a set of $\Nwin$ orthogonal DPSS (see \S \ref{sec:slepian}) as data windows for the SPOD. We restrict this set to the first $2\bw-1$ sequences (equation (\ref{eqn:nwin})) to ensure that all data windows are well-concentrated in $[-\df,\df]$. Since the DPSS are orthogonal, the different realizations are independent. By combining segmenting and orthogonal windowing, a total number of $\Nblk\Nwin$ realizations,
\begin{equation}
\hat{\vb{q}}_{j}^{(k,l)}=\sum_{i=0}^{\Nfft-1}s_l[i]\vb{q}_{i}^{(k)}\mathrm{e}^{\frac{\mathrm{i}2\pi}{\Nfft}ij}, \quad k=1,\dots,\Nblk, \quad l=0,\dots,\Nwin-1,
\label{eqn:fft_slep}
\end{equation}
is obtained. Here, $s_l[i]$ is the $i$-th element of the $l$-th DPSS, $\vb{s}_l$, computed from equation (\ref{eqn:slepian}). The first seven DPSS for $\Nfft=4096$ and $\bw=4$ are shown in figure \ref{fig:slepian} and compared to the Hamming window, equation (\ref{eqn:hamm}). The corresponding sample CSD matrix is
\begin{equation}\label{eqn:csd_mw}
\vb{C}_j = \frac{1}{\Nblk\Nwin}\sum_{k=1}^{\Nblk}\sum_{l=1}^{\Nwin} \hat{\vb{q}}_{j}^{(k,l)} \qty(\hat{\vb{q}}_{j}^{(k,l)})^*.
\end{equation} 
Its matrix representation is also given by equation (\ref{eqn:csd_mat}), but for $\Nsample=\Nblk\Nwin$ and $\hat{\vb{Q}}_{j}=[\hat{\vb{q}}_{j}^{(1,1)}, \hat{\vb{q}}_{j}^{(1,2)},\dots,\hat{\vb{q}}_{j}^{(\Nblk-1,\Nwin)},\hat{\vb{q}}_{j}^{(\Nblk,\Nwin)}]$. 

The standard Welch SPOD algorithm is recovered if a single data window is applied to multiple blocks, ($\Nwin=1$, $\Nblk>1$); a multitaper SPOD algorithm if multiple orthogonal windows are applied to a single segment ($\Nwin>1$, $\Nblk=1$); and a hybrid multitaper-Welch SPOD algorithm if blocking and orthogonal windowing are combined ($\Nwin>1$, $\Nblk>1$).

\subsection{Data}\label{sec:data}

\begin{table}[H]
\centering\label{tab:data}
\begingroup
\setlength{\tabcolsep}{3pt}
\begin{tabular}{lccccccccc}
Example 		& Variables & $N_x$ & $N_{y,r}$ & $N_t$ 	& $\Delta t$		\\ \hline
Jet LES\citep{BresEtAl_2018_JFM}		& $p$	& 261 &	58 & $10000$ 	& 0.2		     \\
Cavity PIV\citep{zhang2019spectral}		& $u,v$	& 78 &	28  & $16000$ 	& $6.24\cdot10^{-5}$	     \\
\end{tabular}
\endgroup
\caption{
Database and spectral estimation parameters. \rii{All quantities are non-dimensionalized as described in the text.}}
\label{tab:data}
\end{table}

Two sets of data are taken as examples. The first data set consists of 10000 snapshots of the pressure field from the wall-modeled large-eddy simulation (LES) of a turbulent $M=0.9$ jet by \citet{BresEtAl_2018_JFM}. This data has been studied in detail using SPOD by \citet{SchmidtEtAl_2018_JFM}. The second data set consists of 16000 snapshots of the streamwise and wall-normal velocity fields of the turbulent $M=0.6$ flow over an open cavity acquired using time-resolved particle image velocimetry (TR-PIV) by \citet{zhang2019spectral}. The open cavity has a length-to-depth ratio of $L/D = 6$ and a width-to-depth ratio of $W/D = 3.85$. More details on the experimental setup can be found in \citet{zhang2017identification}, and an analysis of the data, including using SPOD, in \citet{zhang2019spectral}. \rii{All quantities are non-dimensionalized. For the jet LES data, the pressure is non-dimensionalized by twice the dynamic pressure, $\rho_jU_j^2$, where $U_j$ and $\rho_j$ are the jet exit velocity and density, respectively. Time is non-dimensionalized by $D/c_\infty$, where $c_\infty$ is the far-field speed of sound and $D$ the jet diameter. The dimensionless frequency can therefore be interpreted as the Strouhal number. The radial and streamwise coordinates are likewise given as multiples of $D$. The sampling time step $\Delta t$ corresponds to 200 time steps of the original LES. The 16000 snapshots of the cavity PIV data were recorded over a period of $T=1\;\mathrm{s}$. Using this period to non-dimensionalize time conveniently yields $\Delta f=1$ and the non-dimensional frequency corresponds to the frequency in Hz.} \new{To facilitate the large parameter sweeps conducted for demonstration purposes in this study, the spatial resolutions of both databases were reduced by skipping every other grid point in the streamwise and normal directions. This reduction is accounted for in table \ref{tab:data} and enables, in particular, the memory intensive computation of estimates that use a large number of tapers. We will later demonstrate how the combination of multitapering and segmenting can be used to reduce this computational burden for large data.}

\begin{figure}[H]
  \includegraphics[width=1\textwidth,trim=0 3.2cm 0cm 0.4cm,clip]{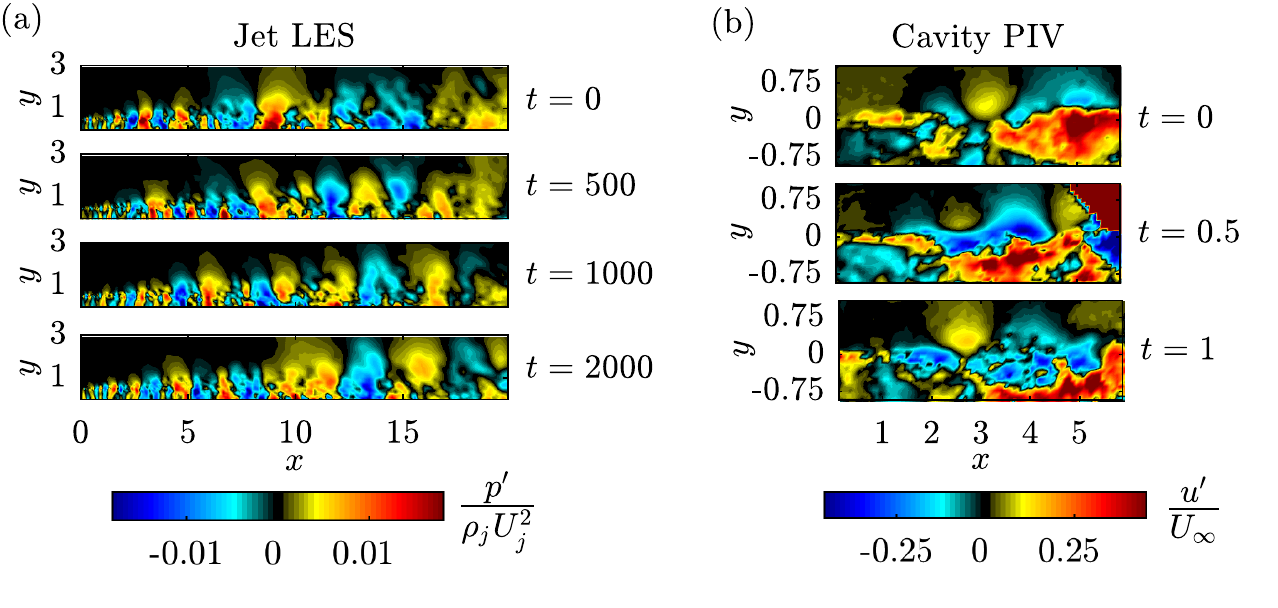}
\caption{Instantaneous flow field visualizations: (a) fluctuating pressure of the turbulent jet LES; (b) streamwise fluctuating velocity of the open cavity PIV data.}
\label{fig:both_instflow}       
\end{figure}
Instantaneous flow field visualizations of both data sets are shown in figure \ref{fig:both_instflow}, and relevant parameters are reported in table \ref{tab:data}. The turbulent jet data are representative of turbulent flows with broadband turbulence spectra \citep{BresEtAl_2018_JFM}. The open cavity data are representative of turbulent flows with tonal peaks and an underlying broadband spectrum \citep{zhang2019spectral}. The two sets also differ in that measurement noise is present only in the experimental cavity flow data.

\section{Results}\label{sec:results}

For the remainder of this paper, we follow best spectral estimation practices \citep{welch1967use}, and set $\Novlp$ to 50\% of $\Nfft$ for all Welch estimates. To compare multiple spectra for different parameters, we first focus on the leading eigenvalues, $\lambda_f^{(1)}$. The variance of the remaining eigenvalues is similar to that of the leading eigenvalue. We do, however, inspect the statistical convergence of the leading two SPOD modes, $\boldsymbol{\phi}_j^{(1)}$ and $\boldsymbol{\phi}_j^{(2)}$, and note that the convergence of the second mode is indicative of the convergence of the remaining modes. The leading SPOD modes play a special role in that they often capture physical mechanisms like resonances or hydrodynamic instabilities that dominate the dynamics at a given frequency. The presence of such mechanisms leads to a rapid convergence of the leading mode and separation of the leading two eigenvalues that represent the mode energies. The latter phenomenon is often referred to as `low-rank behavior', which should be understood in a physical rather than strictly mathematical sense ($\vb{C}_j\in\mathbb{R}^{N\times N}$ is positive semi-definite in general and has full rank only if constructed from $M=N$ independent realizations $\hat{\vb{q}}^{(k)}_{j},\;k=1,\dots,N$). We begin by comparing the basic Welch and multitaper estimates of the SPOD for the open cavity data in \S \ref{sec:results_intro}. In this context, we discuss the bias-variance tradeoff and some subtleties of the estimation process that are best explained by example. A comparative study of both test databases using multitaper-Welch estimates is presented in \S \ref{sec:results_main}.

\subsection{Standard Welch and single-block multitaper estimates of the SPOD}\label{sec:results_intro}

\begin{figure}[H]
  \includegraphics[width=1\textwidth,trim=0 0cm 0cm 0cm,clip]{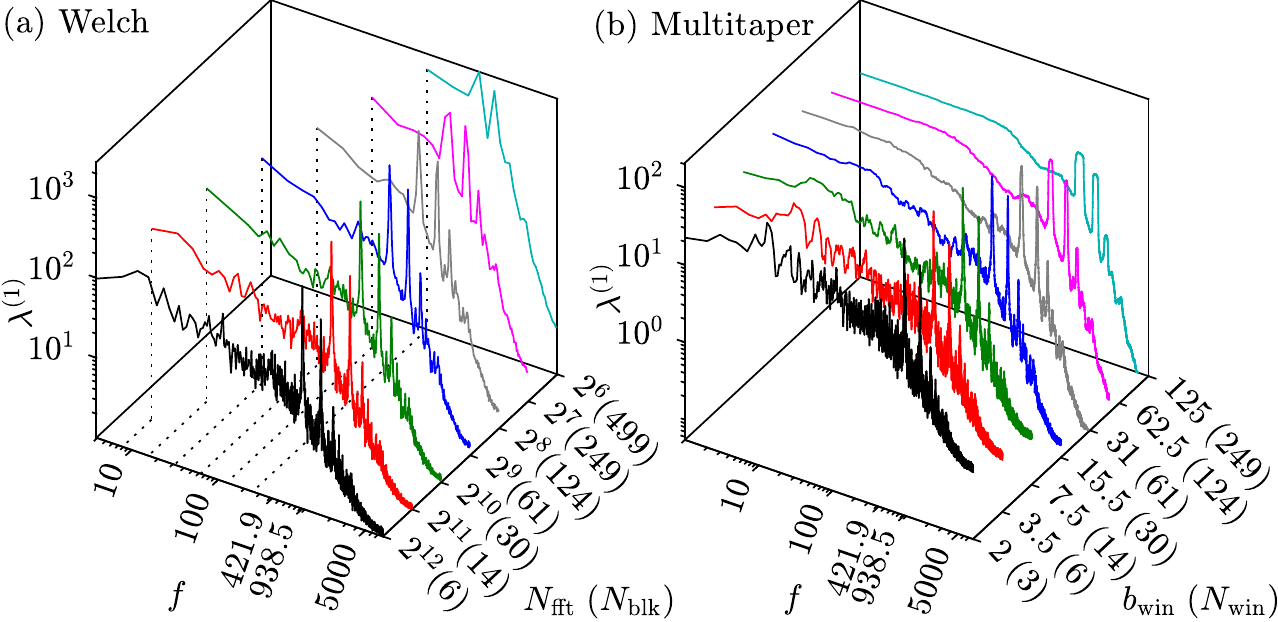}
\caption{Leading SPOD mode spectra of the open cavity PIV data using different estimators: (a) Welch; (b) multitaper. The leading SPOD modes at $f=421.9$ and $f=938.5$ are shown in figure \ref{fig:cavity_modes_comp_WelchVsMultitaper} below. \ri{Dashed lines in (a) indicate the lowest non-zero frequency for each $\Nfft$.}}
\label{fig:WelchVsTaperSpectrum}       
\end{figure}
Figure \ref{fig:WelchVsTaperSpectrum} compares SPOD eigenvalue spectra for the open cavity flow PIV data obtained using standard Welch and single-segment multitaper estimates. The free parameter that is varied for the Welch estimates in figure \ref{fig:WelchVsTaperSpectrum}(a) is the number of samples per block, $\Nfft$. For a fixed overlap of 50\%, this also fixes the number of blocks, $\Nblk$. \ri{For consistency, the Welch estimates are computed with the zeroth DPSS as the data window, and it was confirmed that the Welch estimates of the SPOD using $\vb{s}_0$ and a standard Hamming window, both shown in figure \ref{fig:slepian}, are fundamentally similar.
} The multitaper estimates in figure \ref{fig:WelchVsTaperSpectrum}(b) are obtained for the entire dataset of $\Nfft=N_t=16000$ snapshots, and the time-halfbandwidth product, $\bw$, was varied. The number of tapers used for each estimate, $\Nwin$, then directly follows from equation (\ref{eqn:nwin}). The modes corresponding to two particular frequencies of $f=938.5$ and $f=421.9$ that are labeled on the frequency axis will be discussed later. The higher frequency corresponds to the highest peak in the spectrum and is associated with the first Rossiter mode \citep{rossiter1964wind}, a resonant hydrodynamic instability. The lower frequency corresponds to the much broader peak in the region of high variance that appears close to the Rossiter frequency on the logarithmic scale. 

Welch's method is asymptotically unbiased for stationary data and converges to the true spectral density as $\Nblk,\Nfft\rightarrow\infty$. For limited data, however, care has to be taken regarding the balance of bias and variance. This becomes clear in figure \ref{fig:WelchVsTaperSpectrum}(a). For $\Nfft=2^{12}=4096$ (black line), the data are segmented into only six blocks, and the noisy appearance of the SPOD spectrum indicates the high variance of the estimate. As $\Nfft$ is decreased, the number of blocks increases, and the variance is significantly reduced. Observe, however, that the distinct spectral peaks at the Rossiter and some higher frequencies vanish simultaneously. This is evidence of bias due to the decreased frequency resolution, $\Delta f_\mathrm{Welch} = \frac{1}{\Nfft \Delta t}=\frac{f_s}{\Nfft}$, with decreasing $\Nfft$ and the associated spectral leakage of the under-resolved features. Similar to Welch's method, multitaper estimates of the spectral density based on DPSS are asymptotically unbiased and their mean square error converges to zero as $N\rightarrow\infty$ (but at the same time $\df\rightarrow 0$ with $\bw\rightarrow\infty$, and $\Nwin\leq2\bw$ but $\Nwin\rightarrow\infty$; see \citep{lii2008prolate} for details). As $\bw$ is increased in figure \ref{fig:WelchVsTaperSpectrum}(b), the same qualitative trends are observed as for the Welch estimator, i.e., the variance decreases as the bias increases. The multitaper estimate retains the same frequency resolution, or bin size, of $\Delta f= \frac{1}{\Delta t N_t}=1$, whereas for the Welch estimate, it increases from $\Delta f_\mathrm{Welch} = \frac{1}{\Delta t \Nfft}=3.9$ for $\Nfft=4096$ to $\Delta f_\mathrm{Welch} = 250$ for $\Nfft=64$. For the multitaper estimator at a fixed $N$, the time-halfbandwidth product, $\bw$, determines the frequency bandwidth, $\Delta f_\mathrm{DPSS}=f_s \df$, over which the DPSS tapers yield an integrated average. Welch and multitaper estimates may be compared if they average over a similar frequency band, this is if $\Delta f_\mathrm{Welch}\approx\Delta f_\mathrm{DPSS}$. 

For the Welch estimate with $\Nfft=512$, for example, the frequency resolution is reduced by a factor of $\frac{N_t}{\Nfft}=31.25$. Because $\Delta f = 1$ (16000 samples recorded at 16 kHz over 1 sec) in the experiment, the same value, $\Delta f_\mathrm{Welch} = 31.25$, is obtained for the frequency bin size. Since $\bw$ determines the resolution bandwidth in multiples of $\Delta f$, we consider $\bw=31$, and may confirm that $\Delta f_\mathrm{DPSS} = 31$ (again, because $\Delta f = 1$). We observe that the Welch estimate for $\Nfft=512$ (blue line in figure \ref{fig:WelchVsTaperSpectrum}(a)) and the multitaper estimate for $\bw=31$ (gray line in figure \ref{fig:WelchVsTaperSpectrum}(b)), are indeed qualitatively similar in terms of their variance and resolution of peaks. If the variance is further reduced (cyan lines in figure \ref{fig:WelchVsTaperSpectrum}(a,b)), then the bias is further increased, and spectral peaks are increasingly under-resolved for the Welch estimator, and flattened and broadened for the multitaper estimator. An advantage of multitaper SPOD for comparable bias and variance is that it resolves lower frequencies down to $\Delta f$. A disadvantage is that it becomes computationally expensive for large $\Nwin$, as it requires parallel processing of $\Nwin$ realizations of the Fourier transform of the full data. \new{The computational cost associated with multitaper estimates is discussed in more detail in \S\ref{sec:performance}.}

\begin{figure}[H]
  \includegraphics[width=1\textwidth,trim=0 0cm 0cm 0cm,clip]{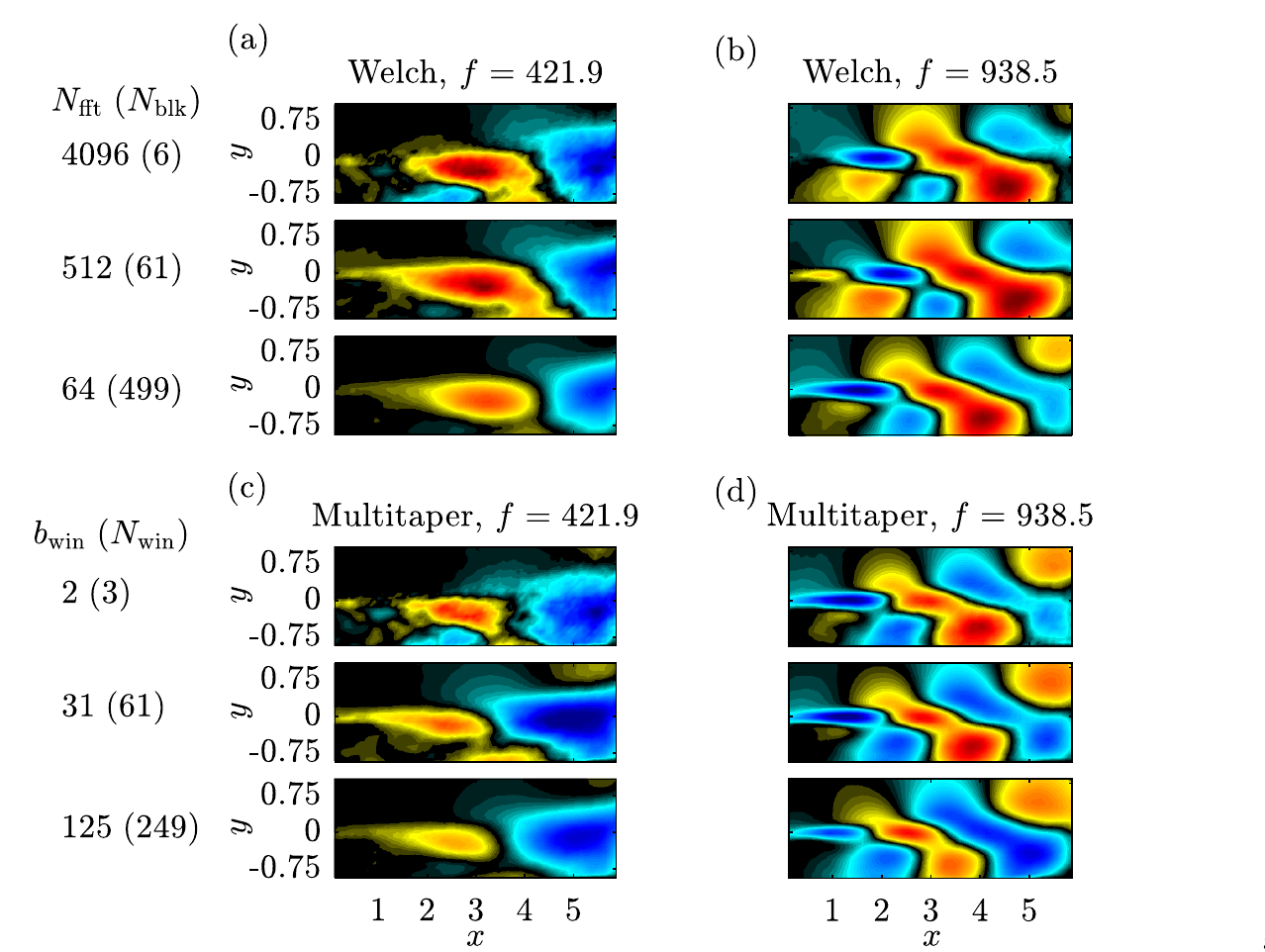}
\caption{Real part of the leading SPOD modes of the open cavity PIV data using different estimators: (a,b) Welch for varying $\Nfft,\Nblk$; (c,d) multitaper for varying $\bw,\Nwin$. The leading SPOD modes at $f=421.9$ and $f=938.5$ are shown in the left and right columns, respectively (labelled in figure \ref{fig:WelchVsTaperSpectrum} above).}
\label{fig:cavity_modes_comp_WelchVsMultitaper}       
\end{figure}

Figure \ref{fig:cavity_modes_comp_WelchVsMultitaper} shows SPOD modes obtained using the Welch (top) and multitaper (bottom) estimators at the two different frequencies, respectively representing the  low-frequency region with high variance (left) and the tonal peak frequency associated with the first Rossiter mode (right). The phases of the complex modes have been approximately aligned for comparability. At $f=938.5$, the Rossiter mode in figure \ref{fig:cavity_modes_comp_WelchVsMultitaper}(b,d) is extracted by both approaches, and well-converged for all parameters. This is typical for hydrodynamic instabilities and resonant mechanisms that are both energetic and evolve coherently over large regions in space and time. At the lower frequency of $f=421.9$, no such mechanism is present, and both the Welch estimate for $\Nblk = 6$ and the multitaper estimate for $\Nwin=3$ appear noisy. Well-converged modes are obtained when $\Nfft$ is decreased (and therefore $\Nblk$ increased), or $\bw$ (and therefore $\Nwin$) are increased for the multitaper estimate. 

\subsection{Multitaper-Welch estimates of the SPOD}\label{sec:results_main}

As demonstrated in \S \ref{sec:results_intro} above, the multitaper method can, in principle,  obtain low-variance estimates without segmentation if a large number of data windows is used. Doing so, however, quickly becomes computationally prohibitive. Furthermore, the estimate at the lowest resolvable frequency relies on a single realization of the flow process if $N=N_t$, and is therefore not statistically representative. These issues are addressed by applying multiple orthogonal windows to overlapping segments, that is, the multitaper-Welch estimate given by equation (\ref{eqn:csd_mw}) with $\Nblk,\Nwin>1$. Our results from \S \ref{sec:results_intro} suggest that Welch and multitaper SPOD yield comparably well-converged modes and spectra for similar bandwidths, $\Delta f_\mathrm{Welch}\approx\Delta f_\mathrm{DPSS}$, leaving computational cost as the only advantage of the Welch estimator. 

\new{For the following study of the performance of the multitaper-Welch estimator}, we chose the highest $\Nfft$ possible that results in the lowest number of blocks that still gives us confidence in the statistical convergence at low frequencies. That segment length is $\Nfft=4096$, resulting in $\Nblk=3$ and $\Nblk=6$ for the jet and cavity data, respectively.

\ri{
\begin{figure}[H]
  \includegraphics[width=1\textwidth,trim=0 0cm 0cm 0cm,clip]{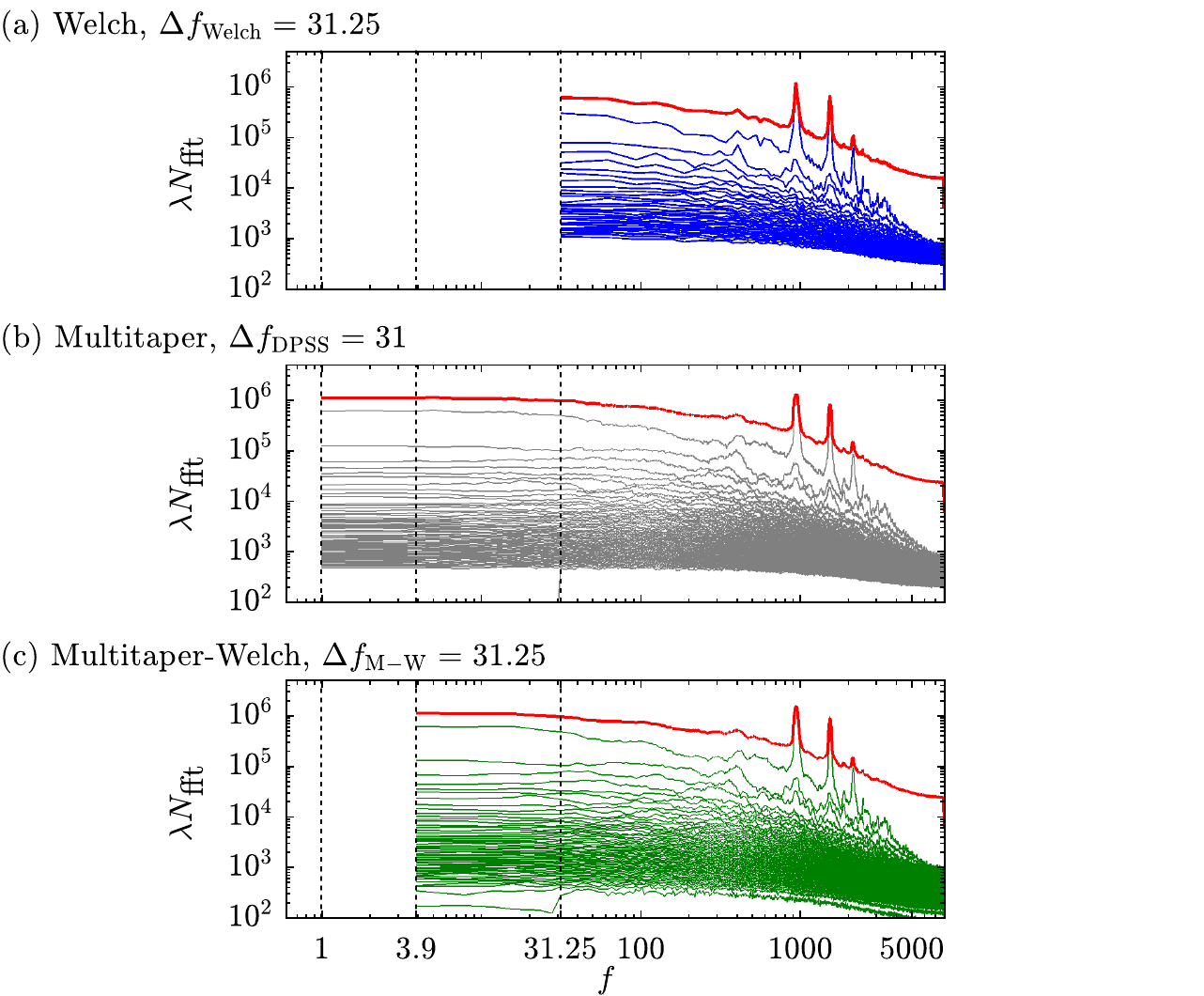}
\caption{\ri{SPOD mode spectra of the cavity PIV data using different estimators with comparable bandwidth, $\Delta f$: (a) Welch with $\Nfft=512$; (b) multitaper with $\bw=31$; (c) multitaper-Welch with $\Nfft=4096, \bw=8$. Colors correspond to those in figures \ref{fig:WelchVsTaperSpectrum}(a), \ref{fig:WelchVsTaperSpectrum}(b), and \ref{fig:both_spectrum}(b) for $\bw=9$, respectively. Red lines show the integral power spectral density (sum of eigenvalues at each frequency) and dotted lines the lowest non-zero frequencies of all three estimates.}}
\label{fig:cavityPIV_fullspec_compBW}       
\end{figure}
Before systematically assessing the effect of the bandwidth parameter, we start with a three-way comparison, including the hybrid multitaper-Welch estimator, at comparable variance. Following the discussion in \S\ref{sec:results_intro}, a comparable variance is achieved by adjusting the spectral estimation parameters such that $\Delta f_\mathrm{Welch}\approx\Delta f_\mathrm{DPSS}\approx\Delta f_\mathrm{M-W}$, where $\Delta f_\mathrm{M-W}$ is the bandwidth of the hybrid estimator. Figure \ref{fig:cavityPIV_fullspec_compBW}(a) and \ref{fig:cavityPIV_fullspec_compBW}(b) show the full SPOD spectra for the Welch and multitaper estimators with $\Delta f\approx 31$, previously shown and discussed in the context of figure \ref{fig:WelchVsTaperSpectrum}(a) and \ref{fig:WelchVsTaperSpectrum}(b). The same colors as in figure \ref{fig:WelchVsTaperSpectrum} are used. Noting that the sum of all eigenvalues corresponds to the energy density, the spectra are multiplied by the number of frequency bins to collapse them on the same scale. The bandwidth of the multitaper-Welch method, $\Delta f_\mathrm{M-W}=\frac{1}{\Nfft \Delta t}\frac{\delta f}{\Delta t}$, results from the combined effects of windowing and tapering. For a fixed segment length of $\Nfft=4096$, we set $\Delta f_\mathrm{M-W}=31.25$ by choosing $\bw=8$. The corresponding SPOD spectrum is shown in green in \ref{fig:cavityPIV_fullspec_compBW}(c) (same color as for the closest bandwidth parameter of $\bw=9$ in figure \ref{fig:both_spectrum} below). We may convince ourselves that the variances of all three estimates in figure \ref{fig:cavityPIV_fullspec_compBW}(a,b,c) are indeed visually very similar. The computational cost of the different estimators is discussed in detail in \S\ref{sec:performance} below. Of the three estimates shown here, the Welch estimate computed that fastest, the Multitaper-Welch estimate took $\sim$15 times, and the Multitaper-only estimate $\sim$45 times as long. The main advantage of the multitaper and hybrid estimators is their higher resolution, in particular at the lower frequencies that contain the largest fraction of the integral energy in most flows. The lowest resolvable frequency for the Welch estimator is $f=31.25$, whereas it is one for the Multitaper method.}

 Next, we vary the bandwidth parameter, $\bw$, and therefore the number of Slepian tapers, $\Nwin$, to systematically study the bandwidth effect.

\begin{figure}[H]
  \includegraphics[width=1\textwidth,trim=0 0cm 0cm 0cm,clip]{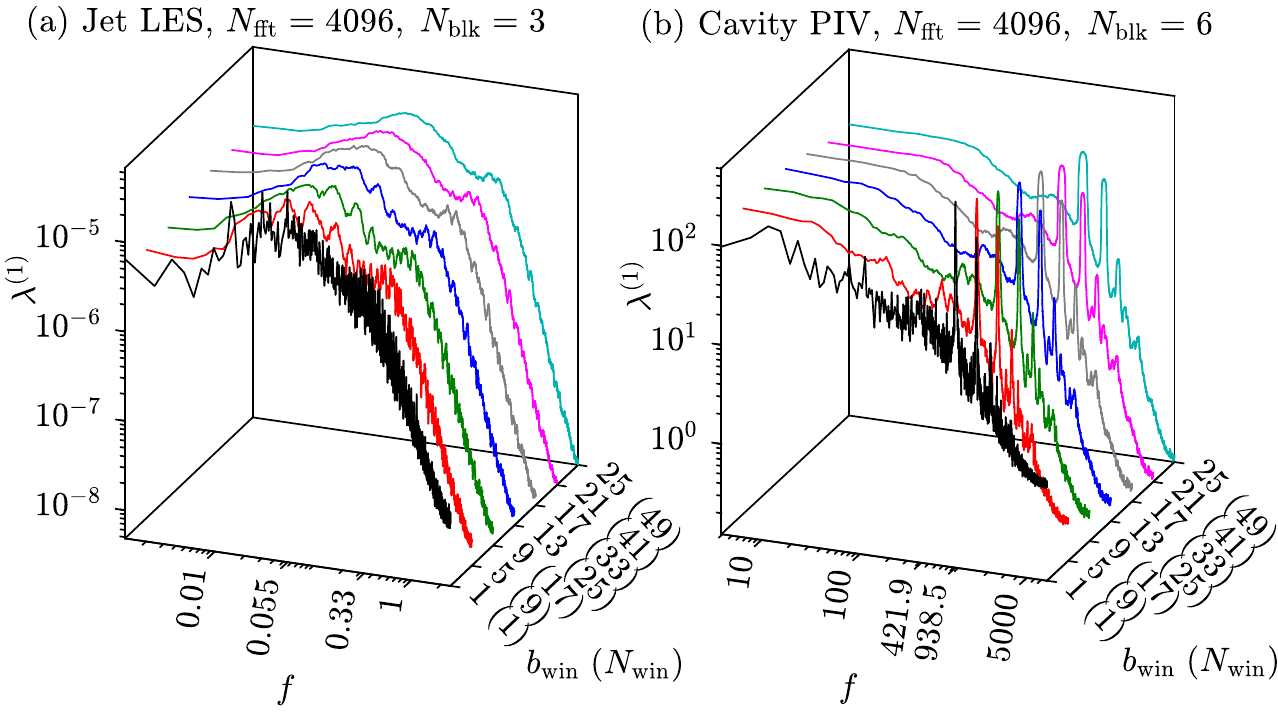}
\caption{Leading SPOD mode spectra using the multitaper-Welch estimator: (a) turbulent jet LES; (b) open cavity PIV data. The leading SPOD modes at $f=0.055$ and $f=0.33$. \new{The segment length is $\Nfft=4096$, resulting in $\Nblk=3$ and $\Nblk=6$ for the jet and cavity data, respectively.}}
\label{fig:both_spectrum}       
\end{figure}
Figure \ref{fig:both_spectrum} shows the SPOD spectra for the leading modes and both cases. The bandwidth parameter is varied from $\bw=1$ to 25, resulting in a number of windows between $\Nwin=1$ (Welch estimator) and 49. The total number of realizations, $\Nblk\Nfft$, is between 3 and 147 for the jet data and between 6 and 294 for the cavity data. A similar decrease in variance is observed for both data as the bandwidth is increased. This decrease is linked to an increase in bias that leads to the loss of certain features, such as several broader peaks at low frequencies in the jet data (still resolved for $\bw=9$, green line) and several double peaks at high frequencies in the cavity data (still clearly visible for $\bw=17$, gray line). Under the caveat that the bias cannot be quantified because the true spectrum is unknown, these observations reflect the usual variance-bias tradeoff.

\begin{figure}[H]
  \includegraphics[width=1\textwidth,trim=0 1.7cm 0cm 0.9cm,clip]{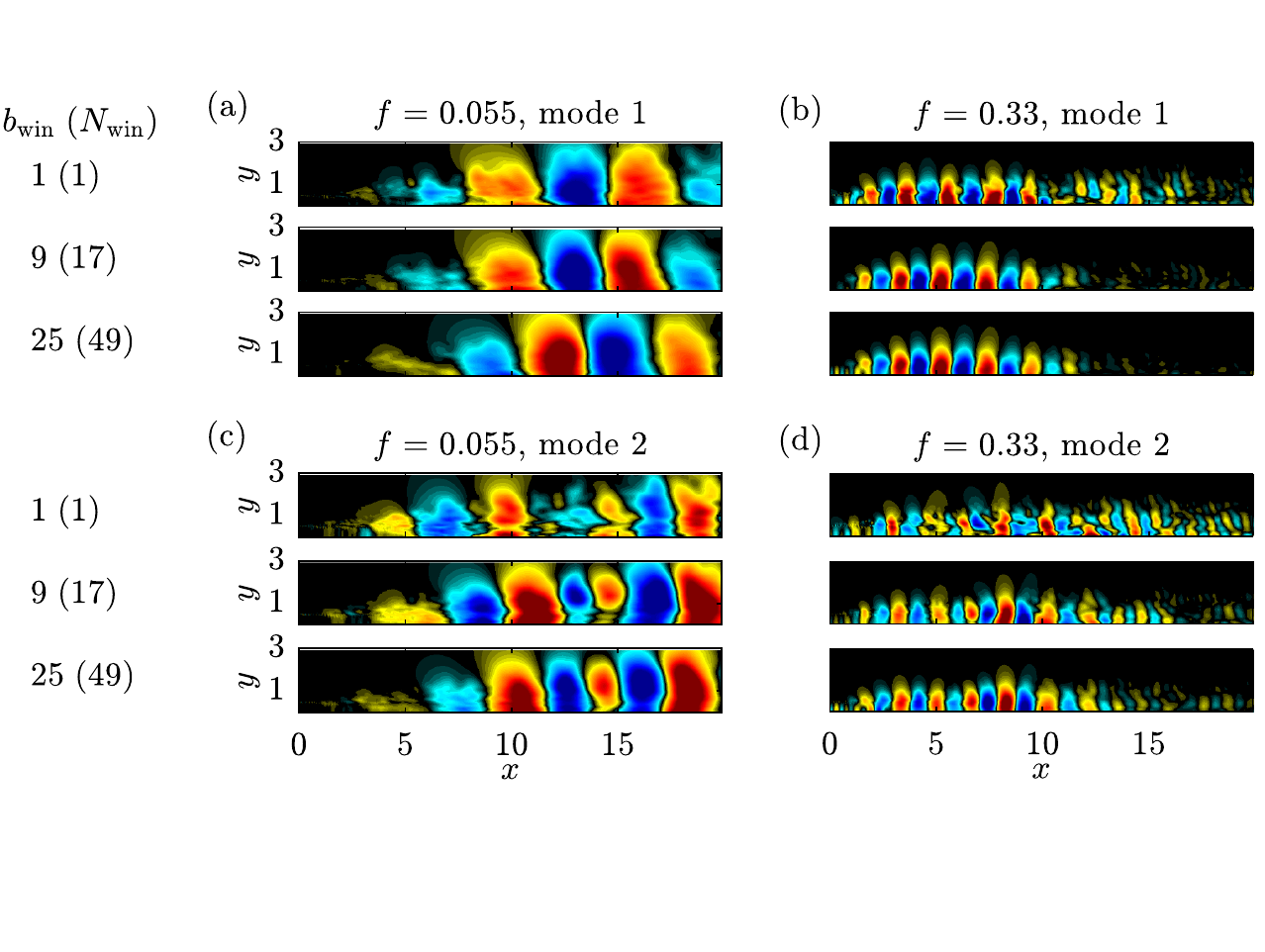}
\caption{Two leading SPOD modes for the jet LES data using multitaper-Welch estimates of different bandwidths: (a) first mode for $f=0.055$; (b) first mode for $f=0.33$; (c) second mode for $f=0.055$; (d) second mode for $f=0.33$. The corresponding leading SPOD energy spectra are reported in figure \ref{fig:both_spectrum}.}
\label{fig:jet_modes_comp}       
\end{figure}
The first and second SPOD modes of the jet data at the two representative frequencies are shown in figure \ref{fig:jet_modes_comp}. Modes obtained for a single, an intermediate, and the highest number of tapers are compared. As expected, better-converged modes are obtained for higher bandwidths. Take as an example the leading SPOD mode for $f=0.33$ shown in \ref{fig:jet_modes_comp}(b). For $\bw=1$ and a single taper, the mode appears noisy, particularly in the downstream region for $x\gtrsim10$. For the two higher bandwidths, the waveform is much more compact and smooth, and very similar for $\bw=9$ and 25. This is indirect evidence of the statistical convergence of the mode. For the second (first subdominant) SPOD mode shown in figure \ref{fig:jet_modes_comp}(d), continual improvement is observed as the bandwidth is increased and a double-lobed structure emerges for the highest value of $\bw=25$.

\begin{figure}[H]
  \includegraphics[width=1\textwidth,trim=0 0cm 0cm 0.4cm,clip]{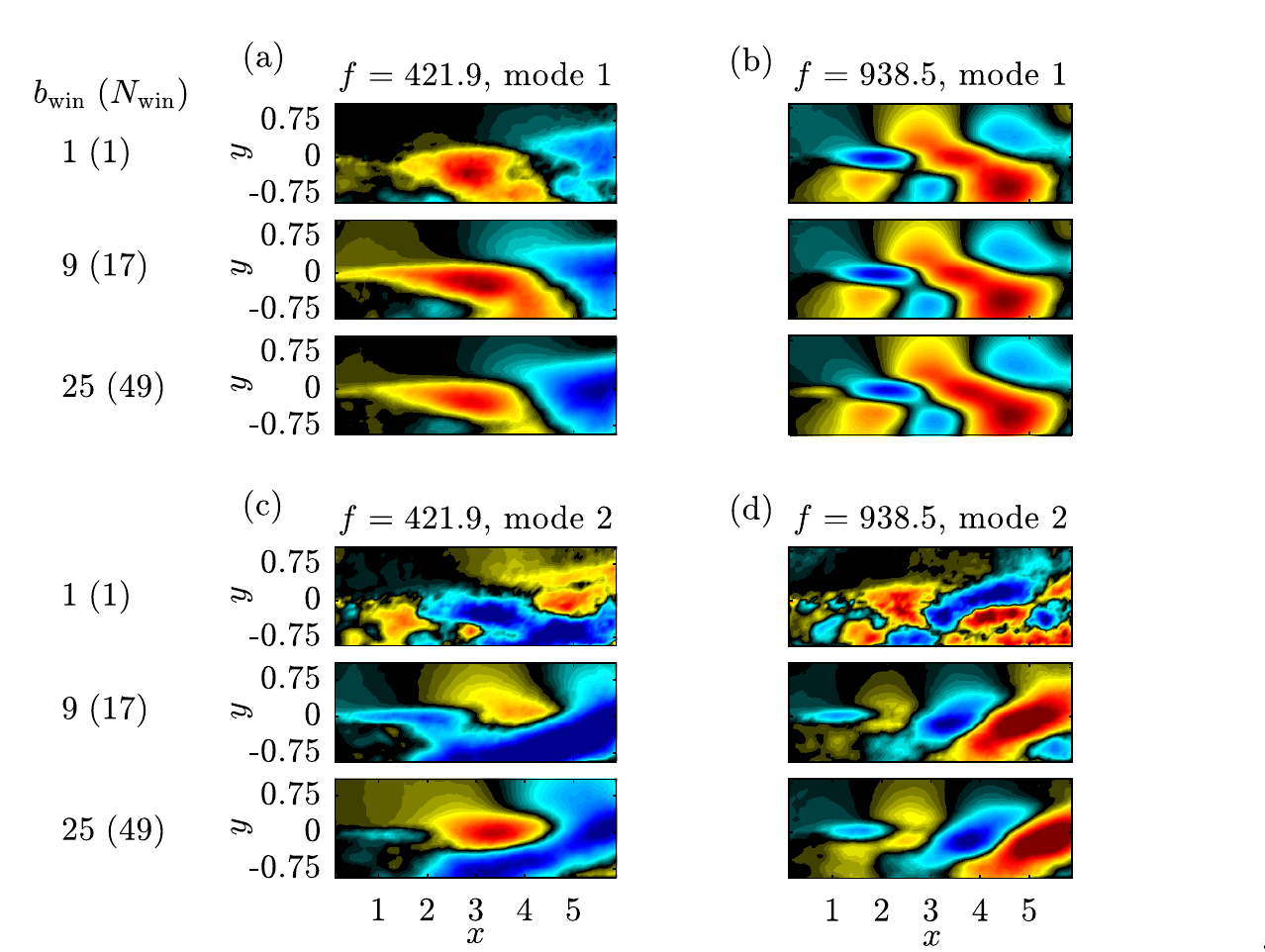}
\caption{Same as figure \ref{fig:jet_modes_comp} but for the cavity PIV data: (a) first mode for $f=421.9$; (b) first mode for $f=938.5$; (c) second mode for $f=421.9$; (d) second mode for $f=938.5$.}
\label{fig:cavity_modes_comp}       
\end{figure}
Similar observations are made for the SPOD modes of the cavity PIV data shown in figure \ref{fig:cavity_modes_comp}. The improvement of mode convergence with increasing bandwidth is apparent, in particular for the second SPOD modes in figure \ref{fig:cavity_modes_comp}(c,d). A clear spatial pattern that is physically interpretable and quantifiable in terms of, for example, streamwise and wall-normal wavenumbers, is only revealed for the two higher bandwidths. A notable exception, as discussed earlier in the context of figure \ref{fig:cavity_modes_comp_WelchVsMultitaper}, is the Rossiter mode shown in figure \ref{fig:cavity_modes_comp}(b) at a frequency of $f=938.5$. It is associated with the overall largest SPOD eigenvalue and is so prevalent that it can be spotted in the instantaneous velocity fields in figure \ref{fig:both_instflow}(b).

\begin{figure}[H]
  \includegraphics[width=1\textwidth,trim=0 0cm 0cm 0cm,clip]{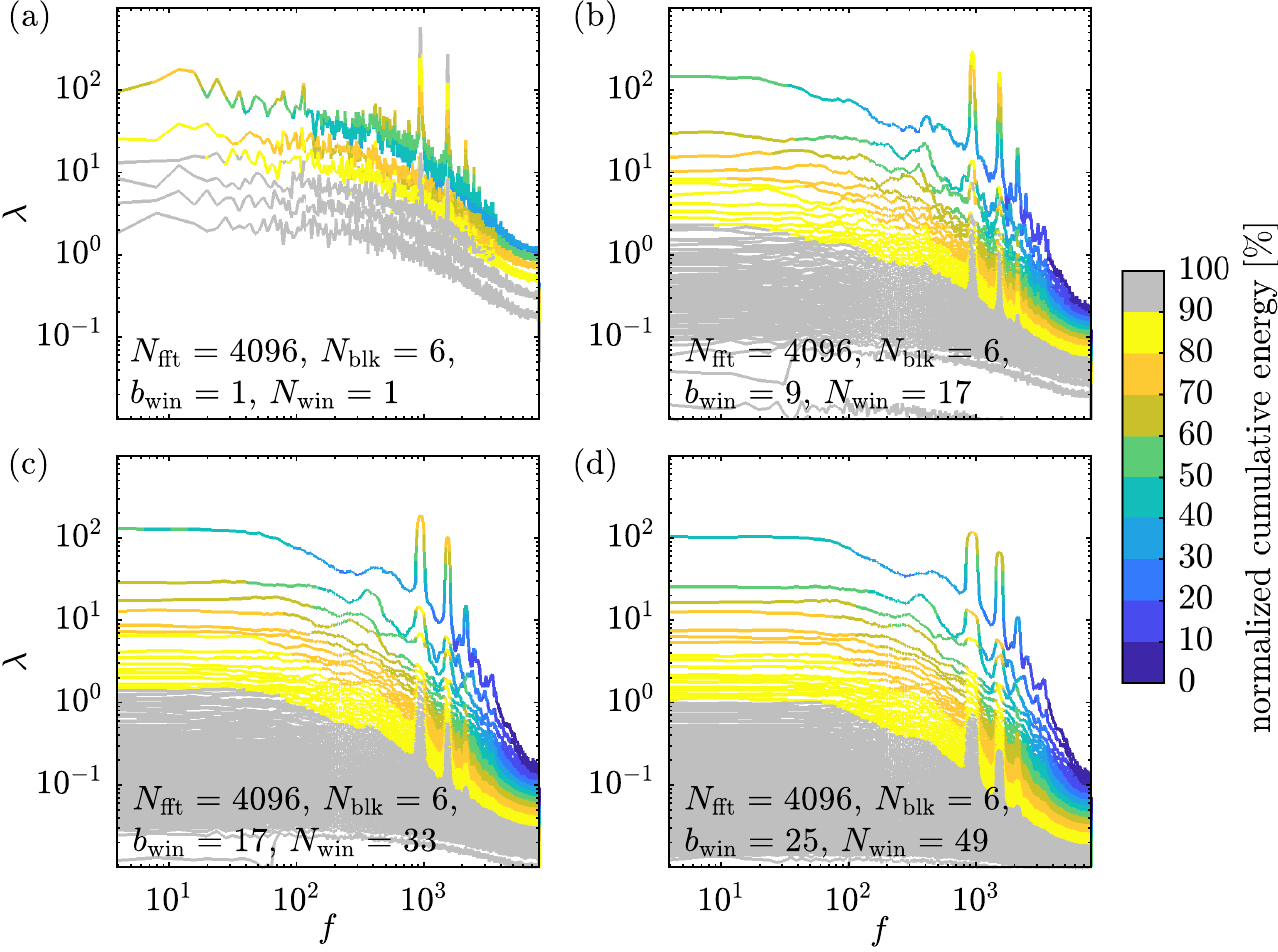}
\caption{Multitaper-Welch SPOD spectra for the cavity PIV data. The data are segmented into $\Nblk=6$ blocks of length $\Nfft=4096$ and $\bw,\Nwin$ are varied: (a) $\bw=1$, $\Nwin=1$; (b) $\bw=9$, $\Nwin=17$; (a) $\bw=17$, $\Nwin=33$; (a) $\bw=25$, $\Nwin=49$.}
\label{fig:cavityPIV_normcumspec_4examples}       
\end{figure}
Finally, we investigate in figure \ref{fig:cavityPIV_normcumspec_4examples} the full SPOD spectrum for the cavity PIV data and the relative energy content of the leading and higher SPOD modes. Four values of $\bw$ between 1, in figure \ref{fig:cavityPIV_normcumspec_4examples}(a), and the maximum of 25, in \ref{fig:cavityPIV_normcumspec_4examples}(d), are considered. It is, as expected, observed that the variance of the higher mode spectra decreases in a very similar manner to that of the leading mode. The normalized cumulative energy, that is, the cumulative sum of the eigenvalues from the highest to the lowest at each frequency, is indicated by the color. The number of modes required to capture 90\% of the energy at any given frequency corresponds to the number of all colored (non-gray) lines. For the two highest peaks and the lowest bandwidth of $\bw=1$ shown in \ref{fig:cavityPIV_normcumspec_4examples}(a), the first mode alone account for more than 90\% of the energy. As the bandwidth is increased, the peaks flatten, and this number reduces to 60\%-80\%. At the same time, the number of modes required to reach 90\% stabilizes, and the main effect of increasing $\bw$ is, apart from the discussed effects on bias and variance, that more eigenvalues of lower energy are added to the bottom of the spectrum. This, again, is an indicator of statistical convergence. 

\section{Computational performance}\label{sec:performance}

\new{
The computational cost of the SPOD algorithm \citep{towneschmidtcolonius_2018_jfm,schmidtcolonius_2020_aiaaj}  implemented in the open-source Matlab code \href{https://www.mathworks.com/matlabcentral/fileexchange/65683-spectral-proper-orthogonal-decomposition-spod}{\tt SPOD}, and extended to multitaper estimators in the present work, is dominated by two main loops (see \citet{SchmidtTowne_2018_CPC} for a detailed description and illustration of the algortihm). The first loop goes over all $\Nblk$ blocks and computes, according to equation \eqn{eqn:fft_slep}, the temporal DFT of each block for each taper. The second loop goes over all $\Nfft$ frequencies. For each frequency, the CSD matrix, as defined in equation \eqn{eqn:C_econ}, is assembled and its eigenvalue decomposition (EVD), equation \eqn{eqn:evp_econ}, is computed. Most of the compute time is, for most cases, spend in this second loop. The only exceptions are the multitaper estimates with the two smallest numbers of tapers ($\Nwin=3$ and $6$), for which the DFTs take most of the time. Clearly, any performance advantage of the multitaper-Welch approach comes at the additional computational cost associated with processing a factor equal to the number of tapers of additional Fourier transforms.
}

\begin{figure}[H]
  \includegraphics[width=1\textwidth,trim=0 0cm 0cm 0cm,clip]{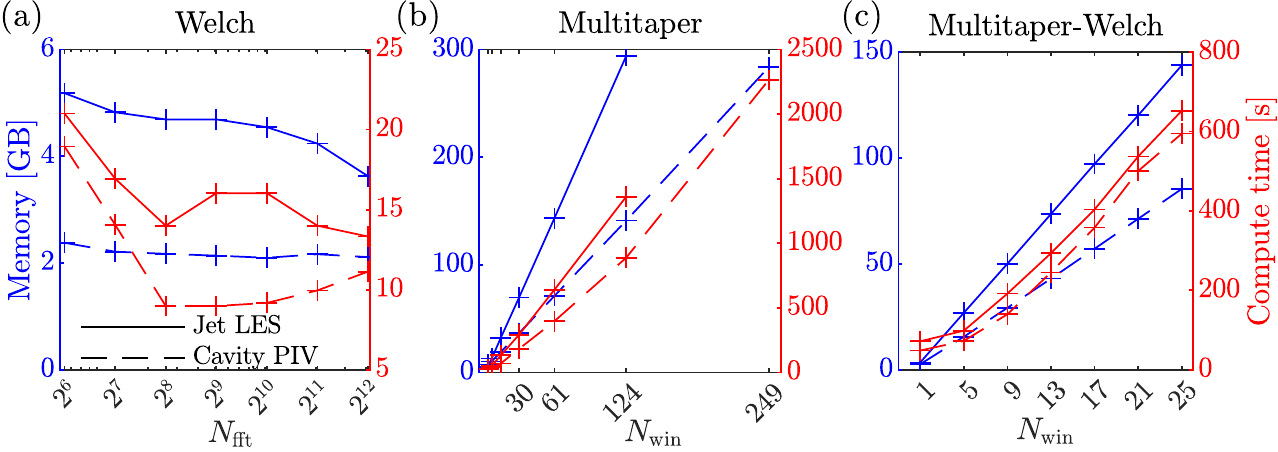}
\caption{\new{Computational performance in terms of memory usage (blue) and compute time (red): (a) Welch; (b) multitaper; (c) multitaper-Welch. Both the jet LES (solid lines) and cavity PIV (dashed lines) data are considered. Refer to figures \ref{fig:WelchVsTaperSpectrum} and \ref{fig:both_spectrum} for the corresponding SPOD spectra. All axes but the abscissa in (a) are on a linear scale.}}
\label{fig:performance}       
\end{figure}

\new{Figure \ref{fig:performance} reports the memory usage and compute times for the Welch and multitaper estimates from \S \ref{sec:results_intro}, and the multitaper-Welch estimates from \S \ref{sec:results_main}. The computations were conducted on a workstation with 20 Intel Xeon cores and 512 GB of memory. The memory consumption and compute times of the Welch estimates shown in figure \ref{fig:performance}(a) vary non-monotonically with the segment length, and do not significantly change as $\Nfft$ is varied from 64 to 4096. This can be explained by the balance between the number of segments and their size that is inherent to the algorithm. If $\Nfft$ is large, then a small number of $\Nblk\Ndof$ DFTs of length $\Nfft$ have to be computed in the first loop, and a large number of $\Nfft$ small EVDs of size $\Nblk\times\Nblk$ have to be solved in the second. These trends are reversed for small $\Nfft$, which results in large $\Nblk$, and the overall computational cost is comparable. A very different behavior is observed for the multitaper and multitaper-Welch estimates in figure \ref{fig:performance}(b) and \ref{fig:performance}(c), respectively. The use of multiple tapers proportionally inflates the number of DFTs that have to be stored in memory and processed from $\Nblk$ to $\Nblk\Nwin$ ($\Nblk=1$ for multitaper-only), independent of their length. This proportional increase of the the problem size with increasing $\Nwin$ is directly reflected in the near-linear scaling of memory and compute time for estimates that use multiple tapers. Note, in particular, that the multitaper-only estimates with the largest values of $\Nwin$ require almost 300 GB of memory for both data sets. The data reduction described in \S\ref{sec:data} was largely motivated by the desire to include these extreme examples in this study for demonstration purposes. As demonstrated in \S\ref{sec:results_intro}, very large values of $\Nwin$ lead to a large bias, and are therefore not recommended in any case. The multitaper-Welch estimates shown and discussed in the context of figure \ref{fig:both_spectrum} provide more balanced results at significantly lower computational cost. The lower computational cost of the hybrid approach becomes apparent from comparing the memory and compute times of the multitaper estimates in panel \ref{fig:performance}(b) to those of the multitaper-Welch algorithm in panel \ref{fig:performance}(c). This assessment of computational performance is taken into account for the best practice recommendations presented next as part of the discussion in \S\ref{sec:discussion}.
}


\section{Discussion}\label{sec:discussion}

In the light of the above comparative study, we start by assessing the performance of multitaper and multitaper-Welch estimates of the SPOD compared to the standard Welch approach. By comparing SPOD spectra obtained using Welch-only and multitaper-only estimators in figure \ref{fig:WelchVsTaperSpectrum}, we have established that both algorithms are comparable in terms of variance, bias, and mode convergence if the frequency resolution of the Welch estimate is similar to the bandwidth of the multitaper estimate. This leaves the multitaper estimate with the advantage of a higher resolution and the associated ability to resolve lower frequencies. Hybrid multitaper-Welch estimates offer the flexibility to control to a large degree the resolution, bias, and variance. In the limit of unit bandwidth and a single taper, the classical Welch estimator is recovered. This allows us to assess the performance advantage of multitaper-Welch estimates over Welch estimates directly from figures \ref{fig:both_spectrum}-\ref{fig:cavityPIV_normcumspec_4examples}, that all contain this limiting case. All comparisons show that additional tapers reduce the variance of the basic Welch estimate and lead to better-converged modes. Similar to the findings of \citet{bronez1992performance} on the estimation of power spectra from time signals, \ri{our results show that the multitaper-Welch approach generally outperforms the Welch-only approach in terms of variance at fixed resolution, and in terms of resolution if the variance is comparable.} This performance advantage comes at the additional computational cost associated with processing a factor equal to the number of tapers of additional Fourier transforms.

We hence recommend using the multitaper-Welch estimator to compute the SPOD and the following best practice: for given data consisting of $N_t$ snapshots, first select $\Nfft$ to obtain the smallest small number of blocks that still ensures that the dynamics at the lowest frequencies are statistically adequately represented. Then choose either the minimum bandwidth necessary to reduce the variance (or increase the mode convergence) to the desired level, \emph{or} select the maximum bandwidth acceptable in terms of bias (or any value in between the two). In settings where well-separated blocks of data are obtained individually, for example, in carefully designed experiments like those by \citet{citriniti2000reconstruction}, multitaper estimates from each block can be leveraged to reduce the number of required blocks, or reduce the variance and improve the mode convergence for an existing set of data.

While SPOD computes modes that optimally represent the data in terms of its second-order statistics, i.e., variance or energy, multitaper estimates are equally applicable to estimate higher-order statistics. They can hence be leveraged to improve estimates of the bispectral mode decomposition \citep{schmidt_2020_nody}. SPOD can quickly become computationally intractable for very large data. This problem can be addressed by updating algorithms that converge the SPOD from streaming data that become available one snapshot at a time. While impossible for the multitaper-only estimator, the algorithm proposed by \citet{SchmidtTowne_2018_CPC} can readily be extended to accommodate multiple data windows and compute multitaper-Welch estimates on the fly.

The multitaper-Welch estimator using DPSS was implemented in the existing open-source Matlab code \href{https://www.mathworks.com/matlabcentral/fileexchange/65683-spectral-proper-orthogonal-decomposition-spod}{\tt SPOD} \citep{towneschmidtcolonius_2018_jfm}. The second argument of the function can be used to specify $\Nfft$ ({\tt NFFT}) and $\bw$ ({\tt BW}) using the syntax {\tt [L,P,F] = SPOD(X,[NFFT BW],...)}. The number of DPSS tapers is then determined by equation (\ref{eqn:nwin}).

\paragraph{Acknowledgments} OTS gratefully acknowledges support from Office of Naval Research grant N00014-20-1-2311 and NSF grant CBET 2046311. I would like to thank Lou Cattafesta and Yang Zhang for providing the TR-PIV data, Gregg Abate for connecting me to Lou, Guillaume Br\`es for generating the turbulent jet database, and Eduardo Martini and Akhil Nekkanti for their constructive feedback. The TR-PIV data was created with support from AFOSR Award Number FA9550-17-1-0380. Creation of the LES data was supported by NAVAIR SBIR under the supervision of J. T. Spyropoulos, with computational resources provided by DoD HPCMP at the ERDC DSRC supercomputer facility.

\paragraph{Data and code availability}
The datasets analyzed in this study are currently not publicly available as they were generated by other researchers (see \S\ref{sec:data}, table \ref{tab:data}) as part of DoD funded independent studies (see \emph{Acknowledgments}), but are available from the corresponding author on reasonable request. A reduced version of the turbulent jet data set is publicly available as part of the open-source Matlab code \href{https://www.mathworks.com/matlabcentral/fileexchange/65683-spectral-proper-orthogonal-decomposition-spod}{\tt SPOD}, that now supports multitaper-Welch estimates for SPOD.

\section*{References}

\bibliography{/Users/oschmidt/Documents/Paper/mypublications_all,/Users/oschmidt/Documents/Paper/DG,/Users/oschmidt/Documents/Paper/PODetc,/Users/oschmidt/Documents/Paper/PSE_and_LST,/Users/oschmidt/Documents/Paper/acoustics,/Users/oschmidt/Documents/Paper/books,/Users/oschmidt/Documents/Paper/cavitation,/Users/oschmidt/Documents/Paper/climate,/Users/oschmidt/Documents/Paper/corner_flow,/Users/oschmidt/Documents/Paper/jets,/Users/oschmidt/Documents/Paper/machine-learning,/Users/oschmidt/Documents/Paper/machineLearning,/Users/oschmidt/Documents/Paper/non-modalStab,/Users/oschmidt/Documents/Paper/others,/Users/oschmidt/Documents/Paper/stochastics,/Users/oschmidt/Documents/Paper/turbulence_and_receptivity,/Users/oschmidt/Documents/Paper/hypersonics,/Users/oschmidt/Documents/Paper/cavity_flow}

\begin{thebibliography}{42}
\providecommand{\natexlab}[1]{#1}
\providecommand{\url}[1]{\texttt{#1}}
\providecommand{\href}[2]{#2}
\providecommand{\path}[1]{#1}
\providecommand{\DOIprefix}{doi:}
\providecommand{\ArXivprefix}{arXiv:}
\providecommand{\URLprefix}{URL: }
\providecommand{\Pubmedprefix}{pmid:}
\providecommand{\doi}[1]{\href{http://dx.doi.org/#1}{\path{#1}}}
\providecommand{\Pubmed}[1]{\href{pmid:#1}{\path{#1}}}
\providecommand{\BIBand}{and}
\providecommand{\bibinfo}[2]{#2}
\ifx\xfnm\undefined \def\xfnm[#1]{\unskip,\space#1}\fi
\makeatletter\def\@biblabel#1{#1.}\makeatother
\bibitem[{Lumley(1970)}]{Lumley:1970}
\bibinfo{author}{Lumley\xfnm[ J.L.]}.
\newblock \bibinfo{title}{Stochastic tools in turbulence}.
\newblock \bibinfo{address}{New York}: \bibinfo{publisher}{Academic Press};
  \bibinfo{year}{1970}.
\bibitem[{Glauser et~al.(1987)Glauser, Leib and George}]{glauser1987coherent}
\bibinfo{author}{Glauser\xfnm[ M.N.]}, \bibinfo{author}{Leib\xfnm[ S.J.]},
  \bibinfo{author}{George\xfnm[ W.K.]}.
\newblock \bibinfo{title}{Coherent structures in the axisymmetric turbulent jet
  mixing layer}.
\newblock \emph{\bibinfo{journal}{Turbulent shear flows}}
  \bibinfo{year}{1987};\bibinfo{volume}{5}:\bibinfo{pages}{134--145}.
\bibitem[{{Schmid}(2010)}]{schmid2010dmd}
\bibinfo{author}{{Schmid}\xfnm[ P.J.]}.
\newblock \bibinfo{title}{{Dynamic mode decomposition of numerical and
  experimental data}}.
\newblock \emph{\bibinfo{journal}{Journal of Fluid Mechanics}}
  \bibinfo{year}{2010};\bibinfo{volume}{656}:\bibinfo{pages}{5--28}.
\newblock \DOIprefix\doi{10.1017/S0022112010001217}.
\bibitem[{McKeon and Sharma(2010)}]{McKeonSharma2010}
\bibinfo{author}{McKeon\xfnm[ B.J.]}, \bibinfo{author}{Sharma\xfnm[ A.S.]}.
\newblock \bibinfo{title}{A critical-layer framework for turbulent pipe flow}.
\newblock \emph{\bibinfo{journal}{Journal of Fluid Mechanics}}
  \bibinfo{year}{2010};\bibinfo{volume}{658}:\bibinfo{pages}{336–382}.
\newblock \DOIprefix\doi{10.1017/S002211201000176X}.
\bibitem[{Towne et~al.(2018)Towne, Schmidt and
  Colonius}]{towneschmidtcolonius_2018_jfm}
\bibinfo{author}{Towne\xfnm[ A.]}, \bibinfo{author}{Schmidt\xfnm[ O.T.]},
  \bibinfo{author}{Colonius\xfnm[ T.]}.
\newblock \bibinfo{title}{Spectral proper orthogonal decomposition and its
  relationship to dynamic mode decomposition and resolvent analysis}.
\newblock \emph{\bibinfo{journal}{Journal of Fluid Mechanics}}
  \bibinfo{year}{2018};\bibinfo{volume}{847}:\bibinfo{pages}{821–867}.
\newblock \DOIprefix\doi{10.1017/jfm.2018.283}.
\bibitem[{Schmidt et~al.(2018)Schmidt, Towne, Rigas, Colonius and
  Br{\`e}s}]{SchmidtEtAl_2018_JFM}
\bibinfo{author}{Schmidt\xfnm[ O.T.]}, \bibinfo{author}{Towne\xfnm[ A.]},
  \bibinfo{author}{Rigas\xfnm[ G.]}, \bibinfo{author}{Colonius\xfnm[ T.]},
  \bibinfo{author}{Br{\`e}s\xfnm[ G.A.]}.
\newblock \bibinfo{title}{Spectral analysis of jet turbulence}.
\newblock \emph{\bibinfo{journal}{Journal of Fluid Mechanics}}
  \bibinfo{year}{2018};\bibinfo{volume}{855}:\bibinfo{pages}{953–982}.
\newblock \DOIprefix\doi{10.1017/jfm.2018.675}.
\bibitem[{Arndt et~al.(1997)Arndt, Long and Glauser}]{arndt1997proper}
\bibinfo{author}{Arndt\xfnm[ R.E.A.]}, \bibinfo{author}{Long\xfnm[ D.F.]},
  \bibinfo{author}{Glauser\xfnm[ M.N.]}.
\newblock \bibinfo{title}{The proper orthogonal decomposition of pressure
  fluctuations surrounding a turbulent jet}.
\newblock \emph{\bibinfo{journal}{Journal of Fluid Mechanics}}
  \bibinfo{year}{1997};\bibinfo{volume}{340}:\bibinfo{pages}{1--33}.
\bibitem[{Citriniti and George(2000{\natexlab{a}})}]{CitrinitiGeorge2000}
\bibinfo{author}{Citriniti\xfnm[ J.H.]}, \bibinfo{author}{George\xfnm[ W.K.]}.
\newblock \bibinfo{title}{Reconstruction of the global velocity field in the
  axisymmetric mixing layer utilizing the proper orthogonal decomposition}.
\newblock \emph{\bibinfo{journal}{Journal of Fluid Mechanics}}
  \bibinfo{year}{2000}{\natexlab{a}};\bibinfo{volume}{418}:\bibinfo{pages}{137--166}.
\bibitem[{Nidhan et~al.(2020)Nidhan, Chongsiripinyo, Schmidt and
  Sarkar}]{nidhanetal_2020_prf}
\bibinfo{author}{Nidhan\xfnm[ S.]}, \bibinfo{author}{Chongsiripinyo\xfnm[ K.]},
  \bibinfo{author}{Schmidt\xfnm[ O.T.]}, \bibinfo{author}{Sarkar\xfnm[ S.]}.
\newblock \bibinfo{title}{{Spectral proper orthogonal decomposition analysis of
  the turbulent wake of a disk at Re=50000}}.
\newblock \emph{\bibinfo{journal}{Physical Review Fluids}}
  \bibinfo{year}{2020};\bibinfo{volume}{5}(\bibinfo{number}{12}):\bibinfo{pages}{124606}.
\newblock \DOIprefix\doi{10.1103/PhysRevFluids.5.124606}.
\bibitem[{Gordeyev and Thomas(2000)}]{gordeyev2000coherent}
\bibinfo{author}{Gordeyev\xfnm[ S.V.]}, \bibinfo{author}{Thomas\xfnm[ F.O.]}.
\newblock \bibinfo{title}{Coherent structure in the turbulent planar jet. part
  1. extraction of proper orthogonal decomposition eigenmodes and their
  self-similarity}.
\newblock \emph{\bibinfo{journal}{Journal of Fluid Mechanics}}
  \bibinfo{year}{2000};\bibinfo{volume}{414}:\bibinfo{pages}{145--194}.
\bibitem[{Gudmundsson and Colonius(2011)}]{gudmundsson2011instability}
\bibinfo{author}{Gudmundsson\xfnm[ K.]}, \bibinfo{author}{Colonius\xfnm[ T.]}.
\newblock \bibinfo{title}{Instability wave models for the near-field
  fluctuations of turbulent jets}.
\newblock \emph{\bibinfo{journal}{Journal of Fluid Mechanics}}
  \bibinfo{year}{2011};\bibinfo{volume}{689}:\bibinfo{pages}{97--128}.
\bibitem[{{Hellstr{\"o}m} and Smits(2014)}]{hellstrom2014}
\bibinfo{author}{{Hellstr{\"o}m}\xfnm[ L.H.O.]}, \bibinfo{author}{Smits\xfnm[
  A.J.]}.
\newblock \bibinfo{title}{The energetic motions in turbulent pipe flow}.
\newblock \emph{\bibinfo{journal}{Physics of Fluids}}
  \bibinfo{year}{2014};\bibinfo{volume}{26}(\bibinfo{number}{12}):\bibinfo{pages}{125102}.
\bibitem[{Tutkun and George(2017)}]{tutkun2017lumley}
\bibinfo{author}{Tutkun\xfnm[ M.]}, \bibinfo{author}{George\xfnm[ W.K.]}.
\newblock \bibinfo{title}{Lumley decomposition of turbulent boundary layer at
  high reynolds numbers}.
\newblock \emph{\bibinfo{journal}{Physics of Fluids}}
  \bibinfo{year}{2017};\bibinfo{volume}{29}(\bibinfo{number}{2}):\bibinfo{pages}{020707}.
\bibitem[{Araya et~al.(2017)Araya, Colonius and Dabiri}]{ArayaEtAl_2017_JFM}
\bibinfo{author}{Araya\xfnm[ D.B.]}, \bibinfo{author}{Colonius\xfnm[ T.]},
  \bibinfo{author}{Dabiri\xfnm[ J.O.]}.
\newblock \bibinfo{title}{Transition to bluff-body dynamics in the wake of
  vertical-axis wind turbines}.
\newblock \emph{\bibinfo{journal}{Journal of Fluid Mechanics}}
  \bibinfo{year}{2017};\bibinfo{volume}{813}:\bibinfo{pages}{346--381}.
\bibitem[{Abreu et~al.(2017)Abreu, Cavalieri and Wolf}]{abreu2017coherent}
\bibinfo{author}{Abreu\xfnm[ L.I.]}, \bibinfo{author}{Cavalieri\xfnm[ A.V.G.]},
  \bibinfo{author}{Wolf\xfnm[ W.]}.
\newblock \bibinfo{title}{Coherent hydrodynamic waves and trailing-edge noise}.
\newblock In: \emph{\bibinfo{booktitle}{23rd AIAA/CEAS aeroacoustics
  conference}}; vol. \bibinfo{volume}{AIAA 2017-3173}.
  \bibinfo{year}{2017}:\unskip\DOIprefix\doi{10.2514/6.2017-3173}.
\bibitem[{He et~al.(2021)He, Fang, Rigas and Vahdati}]{he2021spectral}
\bibinfo{author}{He\xfnm[ X.]}, \bibinfo{author}{Fang\xfnm[ Z.]},
  \bibinfo{author}{Rigas\xfnm[ G.]}, \bibinfo{author}{Vahdati\xfnm[ M.]}.
\newblock \bibinfo{title}{Spectral proper orthogonal decomposition of
  compressor tip leakage flow}.
\newblock \emph{\bibinfo{journal}{Physics of Fluids}}
  \bibinfo{year}{2021};\bibinfo{volume}{33}(\bibinfo{number}{10}):\bibinfo{pages}{105105}.
\bibitem[{Li et~al.(2021)Li, Chen, Liang, Liu and Xiong}]{li2021research}
\bibinfo{author}{Li\xfnm[ X.B.]}, \bibinfo{author}{Chen\xfnm[ G.]},
  \bibinfo{author}{Liang\xfnm[ X.F.]}, \bibinfo{author}{Liu\xfnm[ D.R.]},
  \bibinfo{author}{Xiong\xfnm[ X.H.]}.
\newblock \bibinfo{title}{Research on spectral estimation parameters for
  application of spectral proper orthogonal decomposition in train wake flows}.
\newblock \emph{\bibinfo{journal}{Physics of Fluids}}
  \bibinfo{year}{2021};\bibinfo{volume}{33}(\bibinfo{number}{12}):\bibinfo{pages}{125103}.
\bibitem[{Schmidt et~al.(2019)Schmidt, Mengaldo, Balsamo and
  Wedi}]{schmidtetal_2019_mwr}
\bibinfo{author}{Schmidt\xfnm[ O.T.]}, \bibinfo{author}{Mengaldo\xfnm[ G.]},
  \bibinfo{author}{Balsamo\xfnm[ G.]}, \bibinfo{author}{Wedi\xfnm[ N.P.]}.
\newblock \bibinfo{title}{Spectral empirical orthogonal function analysis of
  weather and climate data}.
\newblock \emph{\bibinfo{journal}{Monthly Weather Review}}
  \bibinfo{year}{2019};\bibinfo{volume}{147}(\bibinfo{number}{8}):\bibinfo{pages}{2979--2995}.
\newblock \URLprefix \url{https://doi.org/10.1175/MWR-D-18-0337.1}.
  \DOIprefix\doi{10.1175/MWR-D-18-0337.1}.
  \href{http://arxiv.org/abs/https://doi.org/10.1175/MWR-D-18-0337.1}{\tt
  arXiv:https://doi.org/10.1175/MWR-D-18-0337.1}.
\bibitem[{Sanjose et~al.(2019)Sanjose, Towne, Jaiswal, Moreau, Lele and
  Mann}]{sanjose2019modal}
\bibinfo{author}{Sanjose\xfnm[ M.]}, \bibinfo{author}{Towne\xfnm[ A.]},
  \bibinfo{author}{Jaiswal\xfnm[ P.]}, \bibinfo{author}{Moreau\xfnm[ S.]},
  \bibinfo{author}{Lele\xfnm[ S.]}, \bibinfo{author}{Mann\xfnm[ A.]}.
\newblock \bibinfo{title}{Modal analysis of the laminar boundary layer
  instability and tonal noise of an airfoil at reynolds number 150,000}.
\newblock \emph{\bibinfo{journal}{International Journal of Aeroacoustics}}
  \bibinfo{year}{2019};\bibinfo{volume}{18}(\bibinfo{number}{2-3}):\bibinfo{pages}{317--350}.
\bibitem[{Nekkanti and Schmidt(2020)}]{nekkantischmidt_2020_aiaaj}
\bibinfo{author}{Nekkanti\xfnm[ A.]}, \bibinfo{author}{Schmidt\xfnm[ O.T.]}.
\newblock \bibinfo{title}{Modal analysis of acoustic directivity in turbulent
  jets}.
\newblock \emph{\bibinfo{journal}{AIAA Journal}}
  \bibinfo{year}{2020};\bibinfo{volume}{0}(\bibinfo{number}{0}):\bibinfo{pages}{1--12}.
\newblock \URLprefix \url{https://doi.org/10.2514/1.J059425}.
  \DOIprefix\doi{10.2514/1.J059425}.
  \href{http://arxiv.org/abs/https://doi.org/10.2514/1.J059425}{\tt
  arXiv:https://doi.org/10.2514/1.J059425}.
\bibitem[{Baars and Tinney(2014)}]{baars2014proper}
\bibinfo{author}{Baars\xfnm[ W.J.]}, \bibinfo{author}{Tinney\xfnm[ C.E.]}.
\newblock \bibinfo{title}{Proper orthogonal decomposition-based spectral
  higher-order stochastic estimation}.
\newblock \emph{\bibinfo{journal}{Physics of Fluids}}
  \bibinfo{year}{2014};\bibinfo{volume}{26}(\bibinfo{number}{5}):\bibinfo{pages}{055112}.
\bibitem[{Nekkanti and Schmidt(2021)}]{nekkantischmidt_jfm_2021}
\bibinfo{author}{Nekkanti\xfnm[ A.]}, \bibinfo{author}{Schmidt\xfnm[ O.T.]}.
\newblock \bibinfo{title}{Frequency–time analysis, low-rank reconstruction
  and denoising of turbulent flows using spod}.
\newblock \emph{\bibinfo{journal}{Journal of Fluid Mechanics}}
  \bibinfo{year}{2021};\bibinfo{volume}{926}:\bibinfo{pages}{A26}.
\newblock \DOIprefix\doi{10.1017/jfm.2021.681}.
\bibitem[{Ghate et~al.(2020)Ghate, Towne and Lele}]{ghate_towne_lele_2020}
\bibinfo{author}{Ghate\xfnm[ A.S.]}, \bibinfo{author}{Towne\xfnm[ A.]},
  \bibinfo{author}{Lele\xfnm[ S.K.]}.
\newblock \bibinfo{title}{Broadband reconstruction of inhomogeneous turbulence
  using spectral proper orthogonal decomposition and gabor modes}.
\newblock \emph{\bibinfo{journal}{Journal of Fluid Mechanics}}
  \bibinfo{year}{2020};\bibinfo{volume}{888}:\bibinfo{pages}{R1}.
\newblock \DOIprefix\doi{10.1017/jfm.2020.78}.
\bibitem[{Chu and Schmidt(2021)}]{chuschmidt_2021_tcfd}
\bibinfo{author}{Chu\xfnm[ T.]}, \bibinfo{author}{Schmidt\xfnm[ O.T.]}.
\newblock \bibinfo{title}{A stochastic spod-galerkin model for broadband
  turbulent flows}.
\newblock \emph{\bibinfo{journal}{Theoretical and Computational Fluid
  Dynamics}} \bibinfo{year}{2021};\URLprefix
  \url{https://doi.org/10.1007/s00162-021-00588-6}.
  \DOIprefix\doi{10.1007/s00162-021-00588-6}.
\bibitem[{Towne(2021)}]{towne2021space}
\bibinfo{author}{Towne\xfnm[ A.]}.
\newblock \bibinfo{title}{Space-time galerkin projection via spectral proper
  orthogonal decomposition and resolvent modes}.
\newblock In: \emph{\bibinfo{booktitle}{AIAA Scitech 2021}}; vol.
  \bibinfo{volume}{AIAA 2021-1676}. \bibinfo{year}{2021}:\unskip
  \bibinfo{pages}{1676}.
\bibitem[{Pain et~al.(2019)Pain, Weiss, Deck and Robinet}]{pain2019large}
\bibinfo{author}{Pain\xfnm[ R.]}, \bibinfo{author}{Weiss\xfnm[ P.E.]},
  \bibinfo{author}{Deck\xfnm[ S.]}, \bibinfo{author}{Robinet\xfnm[ J.C.]}.
\newblock \bibinfo{title}{Large scale dynamics of a high reynolds number
  axisymmetric separating/reattaching flow}.
\newblock \emph{\bibinfo{journal}{Physics of Fluids}}
  \bibinfo{year}{2019};\bibinfo{volume}{31}(\bibinfo{number}{12}):\bibinfo{pages}{125119}.
\bibitem[{Welch(1967)}]{welch1967use}
\bibinfo{author}{Welch\xfnm[ P.]}.
\newblock \bibinfo{title}{{The use of fast Fourier transform for the estimation
  of power spectra: a method based on time averaging over short, modified
  periodograms}}.
\newblock \emph{\bibinfo{journal}{IEEE Transactions on audio and
  electroacoustics}}
  \bibinfo{year}{1967};\bibinfo{volume}{15}(\bibinfo{number}{2}):\bibinfo{pages}{70--73}.
\bibitem[{Schmidt and Colonius(2020)}]{schmidtcolonius_2020_aiaaj}
\bibinfo{author}{Schmidt\xfnm[ O.T.]}, \bibinfo{author}{Colonius\xfnm[ T.]}.
\newblock \bibinfo{title}{Guide to spectral proper orthogonal decomposition}.
\newblock \emph{\bibinfo{journal}{AIAA Journal}}
  \bibinfo{year}{2020};\bibinfo{volume}{58}(\bibinfo{number}{3}):\bibinfo{pages}{1023--1033}.
\newblock \DOIprefix\doi{10.2514/1.J058809}.
\bibitem[{Thomson(1982)}]{thomson1982spectrum}
\bibinfo{author}{Thomson\xfnm[ D.J.]}.
\newblock \bibinfo{title}{Spectrum estimation and harmonic analysis}.
\newblock \emph{\bibinfo{journal}{Proceedings of the IEEE}}
  \bibinfo{year}{1982};\bibinfo{volume}{70}(\bibinfo{number}{9}):\bibinfo{pages}{1055--1096}.
\bibitem[{Bronez(1992)}]{bronez1992performance}
\bibinfo{author}{Bronez\xfnm[ T.P.]}.
\newblock \bibinfo{title}{On the performance advantage of multitaper spectral
  analysis}.
\newblock \emph{\bibinfo{journal}{IEEE Transactions on Signal Processing}}
  \bibinfo{year}{1992};\bibinfo{volume}{40}(\bibinfo{number}{12}):\bibinfo{pages}{2941--2946}.
\bibitem[{Geoga et~al.(2018)Geoga, Haley, Siegel and
  Anitescu}]{geoga_haley_siegel_anitescu_2018}
\bibinfo{author}{Geoga\xfnm[ C.J.]}, \bibinfo{author}{Haley\xfnm[ C.L.]},
  \bibinfo{author}{Siegel\xfnm[ A.R.]}, \bibinfo{author}{Anitescu\xfnm[ M.]}.
\newblock \bibinfo{title}{Frequency–wavenumber spectral analysis of
  spatio-temporal flows}.
\newblock \emph{\bibinfo{journal}{Journal of Fluid Mechanics}}
  \bibinfo{year}{2018};\bibinfo{volume}{848}:\bibinfo{pages}{545–559}.
\newblock \DOIprefix\doi{10.1017/jfm.2018.366}.
\bibitem[{Slepian and Pollak(1961)}]{slepian1961prolate}
\bibinfo{author}{Slepian\xfnm[ D.]}, \bibinfo{author}{Pollak\xfnm[ H.O.]}.
\newblock \bibinfo{title}{Prolate spheroidal wave functions, fourier analysis
  and uncertainty—i}.
\newblock \emph{\bibinfo{journal}{Bell System Technical Journal}}
  \bibinfo{year}{1961};\bibinfo{volume}{40}(\bibinfo{number}{1}):\bibinfo{pages}{43--63}.
\bibitem[{Slepian(1978)}]{slepian1978prolate}
\bibinfo{author}{Slepian\xfnm[ D.]}.
\newblock \bibinfo{title}{Prolate spheroidal wave functions, fourier analysis,
  and uncertainty—v: The discrete case}.
\newblock \emph{\bibinfo{journal}{Bell System Technical Journal}}
  \bibinfo{year}{1978};\bibinfo{volume}{57}(\bibinfo{number}{5}):\bibinfo{pages}{1371--1430}.
\bibitem[{Heinzel et~al.(2002)Heinzel, R{\"u}diger and
  Schilling}]{heinzel2002spectrum}
\bibinfo{author}{Heinzel\xfnm[ G.]}, \bibinfo{author}{R{\"u}diger\xfnm[ A.]},
  \bibinfo{author}{Schilling\xfnm[ R.]}.
\newblock \bibinfo{title}{Spectrum and spectral density estimation by the
  discrete fourier transform (dft), including a comprehensive list of window
  functions and some new at-top windows} \bibinfo{year}{2002};.
\bibitem[{Br{\`e}s et~al.(2018)Br{\`e}s, Jordan, Le~Rallic, Jaunet, Cavalieri,
  Towne, Lele, Colonius and Schmidt}]{BresEtAl_2018_JFM}
\bibinfo{author}{Br{\`e}s\xfnm[ G.]}, \bibinfo{author}{Jordan\xfnm[ P.]},
  \bibinfo{author}{Le~Rallic\xfnm[ M.]}, \bibinfo{author}{Jaunet\xfnm[ V.]},
  \bibinfo{author}{Cavalieri\xfnm[ A.V.G.]}, \bibinfo{author}{Towne\xfnm[ A.]},
  \bibinfo{author}{Lele\xfnm[ S.]}, \bibinfo{author}{Colonius\xfnm[ T.]},
  \bibinfo{author}{Schmidt\xfnm[ O.T.]}.
\newblock \bibinfo{title}{Importance of the nozzle-exit boundary-layer state in
  subsonic turbulent jets}.
\newblock \emph{\bibinfo{journal}{Journal of Fluid Mechanics}}
  \bibinfo{year}{2018};\bibinfo{volume}{851}:\bibinfo{pages}{83–124}.
\newblock \DOIprefix\doi{10.1017/jfm.2018.476}.
\bibitem[{Zhang et~al.(2019)Zhang, Cattafesta and Ukeiley}]{zhang2019spectral}
\bibinfo{author}{Zhang\xfnm[ Y.]}, \bibinfo{author}{Cattafesta\xfnm[ L.]},
  \bibinfo{author}{Ukeiley\xfnm[ L.]}.
\newblock \bibinfo{title}{A spectral analysis modal method applied to cavity
  flow oscillations}.
\newblock In: \emph{\bibinfo{booktitle}{TSFP11, Southampton, UK}}.
  \bibinfo{year}{2019}:\unskip.
\bibitem[{Zhang et~al.(2017)Zhang, Cattafesta and
  Ukeiley}]{zhang2017identification}
\bibinfo{author}{Zhang\xfnm[ Y.]}, \bibinfo{author}{Cattafesta\xfnm[ L.]},
  \bibinfo{author}{Ukeiley\xfnm[ L.]}.
\newblock \bibinfo{title}{Identification of coherent structures in cavity flows
  using stochastic estimation and dynamic mode decomposition}.
\newblock In: \emph{\bibinfo{booktitle}{TSFP10, Chicago, USA}}.
  \bibinfo{year}{2017}:\unskip~\bibinfo{pages}{3}.
\bibitem[{Rossiter(1964)}]{rossiter1964wind}
\bibinfo{author}{Rossiter\xfnm[ J.E.]}.
\newblock \bibinfo{title}{Wind tunnel experiments on the flow over rectangular
  cavities at subsonic and transonic speeds}.
\newblock \emph{\bibinfo{journal}{RAE Technical Report No 64037}}
  \bibinfo{year}{1964};.
\bibitem[{Lii and Rosenblatt(2008)}]{lii2008prolate}
\bibinfo{author}{Lii\xfnm[ K.S.]}, \bibinfo{author}{Rosenblatt\xfnm[ M.]}.
\newblock \bibinfo{title}{Prolate spheroidal spectral estimates}.
\newblock \emph{\bibinfo{journal}{Statistics \& Probability Letters}}
  \bibinfo{year}{2008};\bibinfo{volume}{78}(\bibinfo{number}{11}):\bibinfo{pages}{1339--1348}.
\bibitem[{Schmidt and Towne(2018)}]{SchmidtTowne_2018_CPC}
\bibinfo{author}{Schmidt\xfnm[ O.T.]}, \bibinfo{author}{Towne\xfnm[ A.]}.
\newblock \bibinfo{title}{An efficient streaming algorithm for spectral proper
  orthogonal decomposition}.
\newblock \emph{\bibinfo{journal}{Computer Physics Communications}}
  \bibinfo{year}{2018};\URLprefix
  \url{http://www.sciencedirect.com/science/article/pii/S0010465518304016}.
  \DOIprefix\doi{https://doi.org/10.1016/j.cpc.2018.11.009}.
\bibitem[{Citriniti and
  George(2000{\natexlab{b}})}]{citriniti2000reconstruction}
\bibinfo{author}{Citriniti\xfnm[ J.H.]}, \bibinfo{author}{George\xfnm[ W.K.]}.
\newblock \bibinfo{title}{Reconstruction of the global velocity field in the
  axisymmetric mixing layer utilizing the proper orthogonal decomposition}.
\newblock \emph{\bibinfo{journal}{Journal of Fluid Mechanics}}
  \bibinfo{year}{2000}{\natexlab{b}};\bibinfo{volume}{418}:\bibinfo{pages}{137--166}.
\bibitem[{Schmidt(2020)}]{schmidt_2020_nody}
\bibinfo{author}{Schmidt\xfnm[ O.T.]}.
\newblock \bibinfo{title}{{Bispectral mode decomposition of nonlinear flows}}.
\newblock \emph{\bibinfo{journal}{Nonlinear Dynamics}}
  \bibinfo{year}{2020};(\bibinfo{number}{102(4)}):\bibinfo{pages}{2479--2501}.
\newblock \DOIprefix\doi{10.1007/s11071-020-06037-z}.

\end{thebibliography}

\end{document}